\documentclass[3p,times]{elsarticle}
\pdfoutput=1

\usepackage{amsfonts,amsmath,bm}
\usepackage{algorithmic}
\usepackage{algorithm}
\usepackage{color}
\usepackage{hyperref}

\newtheorem{theorem}{Theorem}
\newtheorem{alg}{Algorithm}
\newtheorem{prop}{Proposition}

\newcommand{\CC}{{\mathbb{C}}}
\newcommand{\Dov}{{D_\text{ov}}}
\newcommand{\kmax}{{k_\text{max}}}
\newcommand{\Ps}{{\tilde P}}

\DeclareMathOperator{\sign}{sign}
\DeclareMathOperator{\spann}{span}
\DeclareMathOperator{\range}{range}
\DeclareMathOperator{\diag}{diag}
\DeclareMathOperator{\re}{Re}

\begin{document}

\title{Short-recurrence Krylov subspace methods for the overlap Dirac
  operator at nonzero chemical potential\tnoteref{t1}}

\author[reg]{Jacques C. R. Bloch}
\author[reg]{Tobias Breu}
\author[wup]{Andreas Frommer}
\author[reg]{Simon Heybrock}
\author[wup]{Katrin Sch\"afer}
\author[reg]{Tilo Wettig}
\address[reg]{Institute for Theoretical Physics, University of
  Regensburg, 93040 Regensburg, Germany}
\address[wup]{Department of Mathematics, University of Wuppertal,
  42097 Wuppertal, Germany}

\tnotetext[t1]{Supported by DFG collaborative research center
  SFB/TR-55 ``Hadron Physics from Lattice QCD''.}

\begin{abstract}
  The overlap operator in lattice QCD requires the computation of the
  sign function of a matrix, which is non-Hermitian in the presence of
  a quark chemical potential.  In previous work we introduced an
  Arnoldi-based Krylov subspace approximation, which uses long
  recurrences. Even after the deflation of critical eigenvalues, the
  low efficiency of the method restricts its application to small
  lattices.  Here we propose new short-recurrence methods which
  strongly enhance the efficiency of the computational method.
  Using rational approximations to the sign function we
  introduce two variants, based on the restarted Arnoldi process and
  on the two-sided Lanczos method, respectively, which become very
  efficient when combined with multishift solvers.
  Alternatively, in the variant based on the two-sided Lanczos method the sign function can be evaluated directly.
   We present numerical
  results which compare the efficiencies of a restarted
  Arnoldi-based method and the direct two-sided Lanczos
  approximation for various lattice sizes.
  We also show that our new methods gain substantially when
    combined with deflation.
\end{abstract}

\maketitle

\section{Introduction}

While this paper discusses new numerical methods that are expected to
be useful in a large number of applications, the main motivation for
these new methods comes from quantum chromodynamics (QCD) formulated
on a discrete space-time lattice.  QCD is the fundamental theory of
the strong interaction.  Being a non-Abelian gauge theory, it is
notoriously difficult to deal with.  Lattice QCD is the only
systematic non-perturbative approach to compute observables from the
theory, and it is amenable to numerical simulations.  The main object
relevant for our discussion is the Dirac operator, for which there
exist several formulations that differ on the lattice but are supposed
to give the same continuum limit when the lattice spacing is taken to
zero.  We are focusing on the overlap Dirac operator $\Dov$
\cite{Narayanan:1994gw,Neuberger:1997fp}, which is the cleanest
formulation in terms of lattice chiral symmetry
\cite{Ginsparg:1981bj,Luscher:1998pqa} but very expensive in terms of
the numerical effort it requires.  Trying to improve algorithms
dealing with the overlap operator is an active field of research, and
even small improvements can have an impact on the large-scale lattice
simulations that are being run by the lattice QCD collaborations
worldwide.

The overlap operator is essentially given by the sign function of its
kernel, which we assume is the usual Hermitian Wilson operator
$H_W=\gamma_5 D_W$ (see \cite{Bloch:2006cd} for the notation).  On the
lattice, this operator is represented by a sparse matrix, and on
current production lattices the dimension of this matrix can be as
large as $10^8\sim10^9$.  The main numerical effort lies in the
inversion of the overlap operator, which is done by iterative methods
and requires the repeated application of the sign function of $H_W$ on
a vector.  At zero chemical potential $\mu$, $H_W$ is Hermitian, and
many sophisticated methods have been developed for this case (see,
e.g., \cite{vandenEshof:2002ms}).  However, one can also study QCD at
nonzero quark chemical potential (or, equivalently, density), which is
relevant for many physical systems such as neutron stars, relativistic
heavy ion collisions, or the physics of the early universe.  The
overlap operator has been generalized to this case
\cite{Bloch:2006cd,Bloch:2007xi}.  While the result is formally
similar to the one at $\mu=0$, it is in fact more complicated since
$H_W$ becomes a non-Hermitian matrix, of which we need to compute the sign function.  This case is much less studied
and the focus of the present paper, which is a natural continuation of
earlier work \cite{Bloch:2007aw}. For simplicity we will
  still refer to $H_W=\gamma_5 D_W$ as the ``Hermitian'' Wilson
  operator.

In mathematical terms, we investigate the computation of $f(A) b$,
where $A \in \CC^{n \times n}$ is non-Hermitian and $f$ is a general
function defined on the spectrum of $A$ such that the extension of $f$
to matrix arguments is defined. For a simple definition of matrix
functions we assume that $A$ is diagonalizable and let $A = R \Lambda
R^{-1}$ be the eigendecomposition with $R\in\CC^{n \times n}$
  and diagonal $\Lambda$ containing the eigenvalues $\lambda_1,
\dots, \lambda_n\in\CC$.  Then the matrix evaluation of $f$ is defined
as
\begin{equation}
  f(A) = R f(\Lambda) R^{-1} = R \diag(f(\lambda_1), \dots,
  f(\lambda_n)) R^{-1}\,.  
\end{equation}
Accordingly, if $b = Ry \in \CC^n$ is a vector expressed in terms of
the eigenvectors, then
\begin{equation}\label{matfun_action:eq}
  f(A)b = R f(\Lambda)y\,.
\end{equation}
For a thorough treatment of matrix functions see \cite{higham08}; a
compact overview is given in \cite{frommer06}. The case $f = \sign$
will be of particular interest. We use the standard definition
$\sign{z} = \sign\re(z)$ for $z \in \mathbb{C}$ \cite{higham08},
which in the physics case we are considering was also shown to follow
from the domain-wall formalism \cite{Bloch:2007xi}.
 
If $A$ is large and sparse, $f(A)$ is too costly to compute, whereas
$f(A)b$ can still be obtained in an efficient manner via a Krylov
subspace method.

The foundation for any Krylov subspace method is the computation of an
appropriate basis for the Krylov subspace $K_k(A, b) = \spann\{b, A b,
\dots, A^{k-1}b\}$. For Hermitian matrices an orthonormal basis can be
built with short recurrences using the Lanczos process. For
non-Hermitian matrices the corresponding process, which again computes
an orthonormal basis, is known as the Arnoldi process. It requires
long recurrences and is usually summarized via the Arnoldi relation
\begin{equation} \label{arnoldi_relation:eq}
  A V_k = V_k H_k + h_{k+1,k} v_{k+1} e_k^T\,.
\end{equation}
Here, $V_k = \left[ v_1 | \cdots | v_k\right] \in \CC^{n\times k}$ is
the matrix which contains the computed basis vectors (the Arnoldi
vectors), $H_k = {V_k}^\dagger A V_k$ is the upper Hessenberg matrix
containing the recurrence coefficients $h_{ij}$, and $e_k$ denotes the
$k$-th unit vector of $\CC^k$. $H_k$ being upper Hessenberg reflects the fact
that the computation of the next Arnoldi vector $v_{k+1}$ results in a
long recurrence since the projection of $Av_k$ on all previous vectors
$v_1, \dots, v_k$ has to be subtracted from $Av_k$. Long recurrences
slow down computation and increase storage requirements, and thus
become inefficient or even infeasible if $k$, the dimension of the
Krylov subspace, becomes large. This is the reason why in this paper
we investigate two ways to circumvent this problem for non-Hermitian
matrices, i.e., restarts of the Arnoldi process and the use of the
two-sided Lanczos process. We will consider these two methods in
  combination with a rational function approximation to $f$.  In the
  case of two-sided Lanczos, we will also consider a direct
  evaluation of the function. 

This paper is organized as follows.  In
  Section~\ref{short:sec} we describe the two alternatives just
  mentioned to obtain short recurrences.  In
  Section~\ref{deflation:sec} we address several aspects of the
  important issue of deflation.  Section~\ref{sec:alg} contains the
  descriptions of four different short-recurrence algorithms to
  compute the sign function, all of which use the preferred method of
  LR deflation.  In Section~\ref{sec:rat} we discuss the choice of the
  rational function to approximate the sign function.  Our numerical
  results are presented in Section~\ref{example:sec}, and conclusions
  are drawn in Section~\ref{sec:concl}.

\section{Short recurrences for non-Hermitian
  matrices} \label{short:sec}

For simplicity, we assume $\| b \|= 1$ from now on.  The standard
Krylov subspace approach, introduced in \cite{vdV87} (see also
\cite{higham08}), to obtain approximations to the action $f(A) b$ is
to compute
\begin{equation}\label{bloch_approx}
  f(A) b \approx V_k {V_k}^\dagger  f(A) b = V_k {V_k}^\dagger  f(A) V_k
  e_1 \approx V_k f(H_k) e_1 
\end{equation}
for some suitable $k \ll n$. Here, $e_1$ denotes the first unit
vector. We refer to \cite{Bloch:2007aw,Bloch:2007jg} for a discussion
in the context of the overlap operator at nonzero chemical potential.

The approximation \eqref{bloch_approx} can be viewed as a
  projection approach. The operator $A$ is orthogonally projected onto
  $K_k(A,b)$, the projection being represented by $H_k= {V_k}^\dagger A
  V_k$. We then compute $f(H_k)e_1$, i.e., we evaluate the matrix
  function of $f$ for the projected operator, applied to the projected
  vector $e_1 = {V_k}^\dagger b$.  This result is finally lifted back to
  the larger, original space by multiplication with $V_k$.  The
matrix function $f(H_k)$, where $H_k$ is of small size, can
be evaluated using existing schemes for matrix functions, e.g., by
computing the eigendecomposition of $H_k$ or by using iterative
schemes like, in the case of $f =\sign$, Roberts' iterative scheme
based on Newton's method, see, e.g., \cite{higham08}.

\subsection{Restarting the Arnoldi process}
\label{sec:restart}

To prevent recurrences from becoming too long for (\ref{bloch_approx})
one could, in principle, use a restart procedure. This means that one
stops the Arnoldi process after $\kmax$ iterations.  At this point
we have a, possibly crude, approximation (\ref{bloch_approx}) to
$f(A)b$, and to allow for a restart one now has to express the error
of this approximation anew as the action of a matrix function,
$f_1(A)b_1$, say.  It turns out that this can indeed be done, see
\cite{eiermann06}, at least in theory, with $f_1$ defined as a divided
difference of $f$ with respect to the eigenvalues of $H_{{\kmax}}$
and with $b_1 = v_{\kmax}$, the last Arnoldi vector of the previous
step. Unfortunately, however, this may result in a numerically
unstable process, so that after a few restarts the numerical results
become useless. For details, see \cite{eiermann06}.

An important exception arises when $f$ is a rational function of the
form
\begin{equation} \label{rational_function:eq}
f(t) =  \sum_{i=1}^s \frac{\omega_i }{t - \sigma_i}\,.
\end{equation} 
We then have 
\begin{equation}\label{schaefer_approx}
f(A) b = \sum_{i=1}^s \omega_i x^{(i)}\,,
\end{equation}
where the $x^{(i)}$, $i=1, \dots, s$, are solutions of the $s$ shifted
systems
\begin{equation}\label{multishift_eq}
(A - \sigma_i I_n) x^{(i)} = b
\end{equation}
and $I_n$ is the $n\times n$ unit matrix (we will frequently
  suppress the index on $I$). For $A$ large and
  sparse, these shifted systems cannot be solved efficiently by direct
  methods.  Using the Arnoldi projection approach outlined before,
the current approximation $x_k$ for $f(A)b$ is obtained as
\begin{equation} \label{rat_arnoldi:eq}
  x_k = \sum_{i=1}^s \omega_i x_k^{(i)}\qquad \mbox{with } 
  x_k^{(i)} = V_k (H_k-\sigma_i I_k)^{-1} e_1\,, \quad i=1,\ldots,s\,.
\end{equation}

Note that Krylov subspaces are shift invariant, i.e., $K_k(A,b) =
K_k(A-\sigma_i I,b)$, and that the Arnoldi process applied to
$A-\sigma_i I$ instead of $A$ produces the same set of Arnoldi
vectors, i.e., the same matrices $V_k$ with $H_k$ replaced by the
shifted counterpart $H_k - \sigma_i I$, see
\cite{Parlett.80,Paige.Parlett.vdVorst.95}. This shows that the
vectors $x_k^{(i)}$ in \eqref{rat_arnoldi:eq} are the iterates of the
\emph{full orthogonalization method} FOM, see \cite{saad96}, for the
linear systems
\begin{equation}
  (A-\sigma_i I) x = b\,.  
\end{equation}

A crucial observation is that for any $k$ the individual residuals
$r^{(i)}_k = b - (A-\sigma_i I)x_k^{(i)}$ of the FOM iterates are just
scalar multiples of the Arnoldi vector $v_{k+1}$, see, e.g.,
\cite{frommer03,simoncini03}, i.e.,
\begin{equation} \label{collinearity:eq}
  r^{(i)}_k = \rho_k^{(i)} v_{k+1}\,, \quad i=1,\ldots,s\,,
\end{equation}
with collinearity factors $\rho_k^{(i)} \in \mathbb{C}$.
The error  $e_k = f(A)b  - x_k$ of  the approximation at step  $k$ can
therefore be expressed as
\begin{equation}
   e_k = f_1(A)v_{k+1}\,, \qquad \mbox{where } f_1(t) = \sum_{i=1}^s
   \frac{\omega_i \rho^{(i)}_k}{t-\sigma_i}\,.   
\end{equation}
This allows for a simple restart at step $\kmax$ of the Arnoldi
process, with the new function $f_1$ again being rational with the
same poles as $f$. For this reason, the stability problems that are
usually encountered with restarts for general functions $f$ do not
occur here.  

The restart process just described can also be regarded as performing
restarted FOM for each of the individual systems $(A-\sigma_i I) x =
b, \, i=1,\ldots,s$ (and combining the individual iterates
appropriately), the point being that, even after a restart, we
need only a single Krylov subspace for all $s$ systems, see
\cite{simoncini03}. Restarted FOM is not the only ``multishift''
solver based on a single Krylov subspace to compute approximations to
$f(A)b$ by combining approximate solutions to $(A-\sigma_i I) x = b$.
An important alternative to FOM is to use restarted GMRES for families
of shifted linear systems as presented in \cite{frommer98}.  This
method also relies on the restarted Arnoldi process, but now a
difference has to be made between the seed system, for which ``true''
restarted GMRES is performed, and the other systems, for which a
variant of GMRES is performed which keeps the residuals collinear to
that of the seed system. The convergence analysis in
\cite{frommer98} shows that this approach is justified if $A$ is
positive real (i.e., $\re( x^\dagger A x) > 0$ for all $x \neq 0$)
and all shifts are negative.\footnote{In our case, $A=H_W^2$ is
    positive real if all the eigenvalues of $H_W$ have their angles in
$(-\frac{\pi}{4},\frac{\pi}{4}) \cup (\frac{3\pi}{4},\frac{5\pi}{4})$, which will be true if $\mu$ is sufficiently small. Experimentally, even for larger values of $\mu$, 
we did not encounter any convergence problems in numerical experiments
    \cite{Schaefer:2008phd}.}
Indeed then,
taking as the seed system the one belonging to the shift
which is smallest in modulus, $\sigma_1$ say, the residuals
of all the other systems --- for which we do not perform ``true'' GMRES
--- are smaller in norm than the residual for $\sigma_1$. But for the
first system we do perform true restarted GMRES which is known to
converge under the assumptions made.

\subsection{The two-sided Lanczos process}

Another way to obtain short recurrences when computing a basis
for the Krylov subspaces for non-Hermitian matrices is
to replace the Arnoldi process by the two-sided Lanczos process. The
two-sided Lanczos process builds two \emph{biorthogonal} bases
$v_1,\ldots,v_k$ and $w_1,\ldots,w_k$ for the two Krylov subspaces
$K_k(A,b)$ and $K_k(A^\dagger,\tilde{b})$, respectively. Here,
$\tilde{b}$ is a so-called shadow vector which can be chosen
arbitrarily.  We always chose $\tilde b = b$ motivated by the
  fact that then for $\mu\to 0$ one recovers the standard Lanczos
  method for which the projection on the Krylov subspace (see
  \eqref{bloch_approx_unsym:eq} below) is
  orthogonal.  With $V_k = \left[v_1 | \cdots | v_k\right]$ and
$W_k = \left[w_1 | \cdots | w_k\right]$ we thus have ${V_k}^\dagger W_k =
I_k$, and the resulting recurrences can be summarized as
\begin{align}
  A V_k &= V_k H_k + h_{k+1,k} v_{k+1} e_k^T\,, \label{2sided1:eq} \\
  A^\dagger  W_k &= W_k {H_k}^\dagger  + \bar{h}_{k,k+1} w_{k+1} e_k^T\,,
  \label{2sided2:eq} 
\end{align}
where $H_k=W_k^\dagger A V_k$ is \emph{tridiagonal}. Note that an iteration will now
require two matrix-vector multiplications, one by $A$ and one
by $A^\dagger$.  In principle, the choice of
  $\tilde{b}$ can substantially influence the two-sided Lanczos
  process, which can even break down prematurely or run into numerical
  instabilities. With our choice of $\tilde{b} = b$ such undesirable
  behavior never occurred in our numerical experiments.

The matrix $V_k{W_k}^\dagger$ now represents an \emph{oblique}
projection, and in analogy to \eqref{bloch_approx} we get the
approximations
\begin{equation}\label{bloch_approx_unsym:eq}
  f(A) b \approx V_k {W_k}^\dagger  f(A) b = V_k {W_k}^\dagger  f(A) V_k e_1 \approx V_k f(H_k) e_1.
\end{equation}
A first report on the application of \eqref{bloch_approx_unsym:eq} to
the overlap operator with chemical potential can be found in
\cite{Bloch:2008gh}.

If $f$ is a rational function, see \eqref{rational_function:eq}, the
approximation \eqref{bloch_approx_unsym:eq} can be expressed as
\begin{equation}
f(A)b \approx \sum_{i=1}^s \omega_i x_k^{(i)} \quad \mbox{with }
x_k^{(i)} = V_k (H_k-\sigma_i I)^{-1} e_1\,.
\end{equation}

Since, just as the Arnoldi process, the two-sided Lanczos process
creates the same vectors $v_k, w_k$ if one passes from $A$ to
$A-\sigma_i I$ with the projected matrix $H_k$ passing to $H_k -
\sigma_i I$, this shows that for all $i$ the vectors $x_k^{(i)}$ are
just the BiCG iterates for the systems $(A-\sigma_i I) x = b$. In
other words: If $f$ is a rational function, the approximation
\eqref{bloch_approx_unsym:eq} is equivalent to performing
``multishift'' BiCG, see \cite{frommer03, Jegerlehner:1996pm} (and
recombining the individual iterates $x_k^{(i)}$ as $\sum_{i=1}^s
\omega_i x_k^{(i)}$).  Although no breakdowns were observed in the
numerical experiments of Ref.~\cite{Schaefer:2008phd}, for reasons of
numerical stability one might prefer using the BiCGStab
\cite{VdVorst:1992} or QMR method \cite{freund92} instead of BiCG.
Both also rely on the two-sided Lanczos process, and efficient
multishift versions exist as well, see
\cite{frommer03,Jegerlehner:1996pm,freund93}.

\subsection{Summary and comparison}
\label{sec:list}

To summarize, so far we have presented the following approaches to
developing short-recurrence methods to iteratively approximate
$f(A)b$:
\begin{enumerate}
\item \emph{Methods based on restarted Arnoldi:} 
  \begin{itemize}
  \item[a)] Approximate $f$ by a rational function
    $g$.  Then use multishift restarted FOM or multishift
    restarted GMRES for $g(A)b$.
  \item[b)] Apply the restarted Arnoldi process directly. As discussed at the beginning of
        Section~\ref{sec:restart}, this is not possible in computational practice because 
       of stability problems.
  \end{itemize}
\item \emph{Methods based on two-sided Lanczos:}
  \begin{itemize}
  \item[a)] Approximate $f$ by a rational function $g$. Then use
    multishift BiCG/BiCGStab/QMR for $g(A)b$.
  \item[b)] Use directly the approximation $V_kf(H_k)e_1$ to the 
       oblique projection $V_k{W_k}^\dagger f(A)b$ for any $f$, see
     \eqref{bloch_approx_unsym:eq}. 
  \end{itemize}
\end{enumerate}
The corresponding algorithms will be given in Section~\ref{sec:alg}.

Note that short recurrences, in principle, result in constant work per
iteration. However, for approach 2b) we will have to evaluate $f(H_k)$
for a $k \times k$ matrix $H_k$, and this work will become substantial
if $k$ is large, see Proposition~\ref{prop:2} below.\footnote{For the
  special case of $f=\sign$, this problem is alleviated by a new
  method \cite{Bloch:2009uc} that speeds up the evaluation of
  $f(H_k)$, thereby eliminating the $\mathcal O(k^3)$ term in
  Proposition~\ref{prop:2}.} Also, in approach 2b) we have to store
all vectors $v_1,v_2,\ldots$ which may become prohibitive so that a
two-pass strategy may be mandatory: The two-sided Lanczos process is
run twice. In the first run, $H_k$ is built up, but the vectors $v_k,
w_k$ are discarded. Once $y_k = f(H_k)e_1$ has been computed, the
Lanczos process is run again, and the vectors $v_k$ are combined with
the coefficients from $y_k$ to obtain the final approximation.  Both
of these drawbacks are not present in the other approaches. These,
however, rely on the fact that we must be able to replace the
computation of $f(A)b$ by $g(A)b$ with a sufficiently precise rational
approximation $g$ to $f$.

\section{Deflation} \label{deflation:sec}

In \cite{Bloch:2007aw} two approaches to deflate eigenvectors were
proposed for the Krylov subspace approximation \eqref{bloch_approx}.
These deflation techniques use eigenvalue information, namely Schur
vectors (Schur deflation) or left and right eigenvectors (LR
deflation) corresponding to some ``critical'' eigenvalues.
Critical eigenvalues are those which are close to a singularity of $f$
since, if these are not reflected very precisely in the Krylov
subspace, we get a poor approximation. In case of the sign function
the critical eigenvalues are those close to the imaginary axis.
In this section we describe both deflation methods and
  show how they can be combined with multishifts so that they can be
  used in approaches based on a rational approximation.  We point out
  a serious disadvantage of Schur deflation, leaving LR deflation as
  the method of choice. For the sake of simplicity we present the deflation techniques 
without taking restarts into account. We will briefly comment on restarts
after \eqref{exact_proj:eq} below.

We start with Schur deflation. Let $S_m = [s_1 | \cdots | s_m]$ be the matrix whose columns $s_i$ are
the Schur vectors of $m$ critical eigenvalues of the matrix $A$. This
means that we have ${S_m}^\dagger S_m = I_m$ and
\begin{equation}\label{Schur}
  A S_m = S_m T_m\,,
\end{equation}
where $T_m$ is an upper triangular matrix with the $m$ critical
eigenvalues of $A$ on its diagonal, see \cite{golub}. Let us note that
the Schur vectors span an invariant subspace of $A$, and that they can
be computed via orthogonal transformations, which is very stable
numerically.  The extraction of the eigenvectors themselves is a less
stable process if $A$ is non-Hermitian.

In the case of the shifted matrices $A - \sigma_i I$, $i = 1, \dots,
s$, and $S_m$, $T_m$ computed with respect to $A$ we have
\begin{equation}
  (A - \sigma_i I) S_m = A S_m - \sigma_i S_m = S_m (T_m - \sigma_i
  I_m)\,, \qquad i = 1, \dots, s\,. 
\end{equation} 
Clearly, the matrix $\Ps = S_m {S_m}^\dagger $ represents the orthogonal
projector onto the subspace $\Omega_m = \spann\{s_1, \dots, s_m\}$.
The solutions to \eqref{multishift_eq} are now approximated in 
augmented Krylov subspaces,
\begin{equation}
x_k^{(i)} \in  \Omega_m + (I - \Ps)K_k(A,b)\,.
\end{equation}
In fact, the projected Krylov subspace $(I - \Ps)K_k(A,b)$, which is orthogonal to $\Omega_m$, is a Krylov
subspace again, but now for $(I-\Ps)A$ instead of $A$ and $(I-\Ps)b$
instead of $b$: Since $\Omega_m = \range(\Ps)$ is $A$-invariant, i.e.,
for any $y$ there is a $\tilde{y}$ such that $A \Ps y = \Ps \tilde{y}$, we
have
\begin{equation}
  (I-\Ps)A(I-\Ps)y = (I-\Ps)Ay - (I-\Ps)A\Ps y = (I-\Ps)Ay - (I-\Ps)\Ps\tilde{y} = (I-\Ps)A y
\end{equation}
and thus
\begin{align}\label{projected_subspace}
  (I-\Ps) K_k(A, b) &= \spann\{(I-\Ps)b, (I-\Ps)Ab , \dots, (I-\Ps)A^{k-1}b\}
  \nonumber\\ 
  &= \spann\{(I-\Ps)b, (I-\Ps)A(I-\Ps)b , \dots, ((I-\Ps)A)^{k-1}(I-\Ps)b\}
  \nonumber\\ 
  &= K_k((I-\Ps)A,(I-\Ps)b)\:.
\end{align}
To build a basis $V_k = \left[v_1 | \cdots | v_k\right]$ for this
Krylov subspace we have to multiply by $(I-\Ps)A$ instead of $A$ in
every step, reflecting the fact that we have to project out the
$\Omega_m$-part after every multiplication by $A$. This may result
in quite considerable computational work: The work for one projection
has cost $\mathcal{O}(nm)$, because each of the $m$ Schur vectors is
usually non-sparse.

We now turn to LR deflation. The idea is essentially the same as for Schur deflation, except that we use a different projector. As we will see below, this has a useful consequence: It removes the need to multiply by $I-\Ps$ in every step. Thus the $\mathcal{O}(nm)$ effort for the projection step has to be paid only once, instead of once per iteration (but see the comment after Eq.~\eqref{eq:cont}). 

The projector we use is an oblique projector onto $\Omega_m$, defined by $P = R_m L_m^\dagger $, where $R_m = \left[r_1 | \cdots | r_m\right]$ is
  the matrix containing the right eigenvectors and $L_m^\dagger =
\left[l_1 | \cdots | l_m\right]^\dagger $ is the matrix containing the
left eigenvectors corresponding to the $m$ critical eigenvalues of $A$.
With $\Lambda_m$ the diagonal matrix with the $m$ critical eigenvalues on
its diagonal, the left and right eigenvectors satisfy
\begin{equation}
  A R_m = R_m \Lambda_m \quad \text{and} \quad L_m^\dagger  A =
  \Lambda_m L_m^\dagger \,. 
\end{equation}
The left and right eigenvectors are biorthogonal and are normalized
such that $L_m^\dagger R_m = I_m$, thus ensuring $P^2=P$.

As in the Schur deflation the projected Krylov subspace
$(I-P)K_k(A,b)$ is a Krylov subspace.  It is no longer orthogonal to
$\Omega_m$ because the projector is oblique, but it now is a Krylov
subspace for the original matrix $A$ since both $\range(P)$ and
$\range(I-P)$ are $A$-invariant so that $(I-P) A y = A y$ for $y \in
\range(I-P)$.  Instead of \eqref{projected_subspace} we now have
\begin{equation}
  \label{eq:cont}
  (I-P)K_k(A,b) = K_k(A,(I-P)b)\,.
\end{equation}
Therefore, no additional projection is needed within the Arnoldi
method when we build up a basis $V_k = \left[v_1 | \cdots |
  v_k\right]$ for this subspace.  In computational practice,
  however, components outside of $\range(I-P)$ will show up gradually
  due to rounding effects in floating-point arithmetic. It is thus
  necessary to apply $(I-P)$ from time to time in order to eliminate
  these components. We will come back to this point in
  Section~\ref{LR_rat:sec}.

  The numerical accuracy of the computed \emph{eigenvectors} turned
  out to always be sufficient in our computations. Therefore, because
  of its greater efficiency, from now on we concentrate on LR rather
  than Schur deflation.

The overall approach is thus as follows:
With the oblique projector $P = R_m L_m^\dagger$ we split 
$f(A)b$ into the two parts
\begin{equation}\label{eq:action}
   f(A)b = f(A)(Pb) + f(A)(I-P)b\,.
\end{equation}
Since we know the left and right eigenvectors which make up $P$, using
\eqref{matfun_action:eq} we directly obtain
\begin{equation} \label{exact_proj:eq}
 x_P \equiv f(A)(Pb) = f(A)R_m L_m^\dagger b = R_m f(\Lambda_m) (L_m^\dagger b)\,.
\end{equation}
The remaining part $f(A)(I-P)b$ can then be approximated iteratively by
any of the approaches discussed in Section~\ref{short:sec}.
Since the only effect of LR deflation is the replacement of $b$
  by $(I-P)b$ in \eqref{bloch_approx}, no modifications are necessary
  when using one of the restarted approaches.

There is a beneficial
effect of deflation on the number of poles to use when
$f$ is approximated by a rational function $g$.
Let $y$ be the coefficient vector of $b$ when represented in the basis
of right eigenvectors of $A$, i.e., $b = Ry$, and assume that we
sorted them to put the critical eigenvectors first, i.e.,
\begin{equation}
  R = \left[R_m \left| R_{\neg m} \right. \right]\,, \qquad 
  y = \left[ \begin{array}{c} y_m \\ y_{\neg m} \end{array} \right]\,, 
  \qquad \Lambda = \left[ \begin{array}{cc} \Lambda_m & 0 \\
      0 & \Lambda_{\neg m} \end{array} \right]\,.
\end{equation}
Then $f(A)Pb = R_mf(\Lambda_m)y_m$ and $f(A)(I-P)b = R_{\neg
  m}f(\Lambda_{\neg m})y_{\neg m}$. So when approximating $f(A)(I-P)b$
via a rational function $g$, we have $f(A)(I-P)b \approx g(A)(I-P)b =
R_{\neg m}g(\Lambda_{\neg m})y_{\neg m}$. This shows that we only have
to take care that $g$ approximates $f$ well on the non-critical
eigenvalues (those in $\Lambda_{\neg m}$). Consequently, a good
approximation can be obtained using a smaller number of poles as
compared to the situation where we would have to approximate well on
the full spectrum of $A$. This pole-reduction phenomenon can be very
substantial, even if we deflate only a small number of eigenvalues,
see Section~\ref{example:sec}. 

\section{Algorithms}
\label{sec:alg}

In this section we present four algorithms, corresponding to the list
in Section~\ref{sec:list}, to compute the action \eqref{eq:action} of
the sign function of a non-Hermitian matrix on a vector, using LR
deflation for the first term $f(A)(Pb)$ and short-recurrence
Krylov subspace methods for the remaining term
$f(A)(I-P)b$.

\subsection{Restarted Arnoldi with rational
  functions} \label{LR_rat:sec}

In this subsection we discuss methods based on restarted Arnoldi,
corresponding to 1a) in Section~\ref{sec:list}.  We assume that the
original function $f$ is replaced by a rational function given by
\eqref{rational_function:eq} which approximates the original function
sufficiently well.  The choice of the rational function will be
discussed in Section~\ref{sec:rat}.

We start with LR-deflated restarted FOM. The resulting algorithm is
given as Algorithm~\ref{fom:alg}, where we use \eqref{rat_arnoldi:eq}
to obtain the iterates for all shifted systems in the current cycle,
and where we give the details on how to obtain the collinearity
factors $\rho_k^{(i)}$ from \eqref{collinearity:eq} for the residuals,
see also \cite{simoncini03}. Here, the notation FOM-LR($m,k$)
indicates that we LR-deflate a subspace of dimension $m$ and that we
restart FOM after a cycle of $k$ iterations.  The vector $x$ is the
approximation to $f(A)b$.  After the completion of each cycle
  we perform a projection step to eliminate numerical contamination by
  components outside of $\range(I-P)$, as discussed in
  Section~\ref{deflation:sec} after \eqref{eq:cont}.
 
\begin{algorithm}
\begin{alg}\rm Restarted FOM-LR($m,k$)\label{fom:alg}
\begin{algorithmic}
\STATE \{{\bf Input} $m$, $k=\kmax$, $A$, $\{\sigma_1, \dots, \sigma_s\}$, $\{\omega_1,\ldots,
\omega_s\}$, $b$, $L=L_m$, $R=R_m$, $\Lambda=\Lambda_m$\}
\STATE
\STATE $x = x_P = R f(\Lambda) {L}^\dagger b$
\STATE $r = (I-P)b$
\STATE $\rho^{(i)} = 1$, $i = 1, \dots, s$
\WHILE[\emph{loop over restart cycles}]{not all systems are converged}
\STATE $\beta = \|r\|_2$
\STATE $v_1 = r/\beta$
\STATE compute $V_k$, $H_k$ by running $k$ steps of Arnoldi with $A$
\FOR{$i = 1, \dots, s$}
\STATE $y_k^{(i)} = \beta \rho^{(i)} (H_k - \sigma_i I_k)^{-1} e_1$
\ENDFOR
\STATE $x = x + V_k\sum_{i=1}^s \omega_i y_k^{(i)}$
\STATE $r = v_{k+1}$
\STATE $\rho ^{(i)} = - h_{k+1,k} e_k^T y_k^{(i)}$, $i = 1, \dots, s$
\STATE $r = (I-P)r$ \COMMENT{\emph{projection step}}
\ENDWHILE
\end{algorithmic}
\end{alg}
\end{algorithm} 

Since Algorithm \ref{fom:alg} will be used in our numerical
experiments, we now analyze the main contributions to its
computational cost.

\begin{prop} Let $C_n$ denote the cost for one matrix-vector
  multiplication by the matrix $A$, and let $k_{\rm tot}$ be the total
  number of such matrix-vector multiplications performed. The
  computational cost of Algorithm~\ref{fom:alg} is given as
\begin{equation} \label{cost_fom:eq}
k_{\rm tot} \Big[C_n + n \, \big[\mathcal{O}(k_{\rm max}) +
    \mathcal O(m/k_ {\rm max})\big]\Big]\,.
\end{equation}   
\end{prop}
To see this, let us discuss the dominating contributions to the
computational cost in one sweep of the while-loop. For simplicity we
write $k$ instead of $k_\text{max}$, as we also did in the
algorithm. Computing $V_k$ and $H_k$ with the Arnoldi process has cost
$kC_n + \mathcal{O}(nk^2)$, since for $j=1,\ldots,k$ the $j$-th step
requires one matrix-vector multiplication and $j$ inner products,
vector additions and scalings. Since we can solve systems with the
upper Hessenberg matrices $H_k - \sigma_i I_k$ with cost
$\mathcal{O}(k^2)$, the total cost for the computation of all the
vectors $y_k^{(i)}$ is $\mathcal{O}(s k^2)$, which can be neglected
compared to the $\mathcal{O}(nk^2)$ cost contained in the Arnoldi
process.  Updating $x$ with the linear combination of the columns of
$V_k$ has cost $\mathcal{O}(ks+nk)$, which can again be neglected
compared to the $\mathcal{O}(nk^2)$ cost in the Arnoldi process.  The
final projection step has cost $\mathcal O(mn)$. Multiplying these
costs by the number $n_\text{sweep}$ of sweeps through the while-loop
and using $k_\text{tot} = n_\text{sweep}k_\text{max}$ gives the total
cost. The initial steps prior to the while loop have cost $\mathcal
O(mn)$, which is dominated by the last term of
Eq.~\eqref{cost_fom:eq}.

We now formulate the LR-deflated restarted GMRES algorithm.  Let us first
introduce the $(k+1)\times k$ matrix
\begin{equation}
  \widehat{H}_k = \left[ \begin{array}{c} H_k \\
                                    h_{k+1,k}e_k^T
                     \end{array} 
               \right]
\end{equation}
through which the Arnoldi relation \eqref{arnoldi_relation:eq} can be
summarized as $AV_k = V_{k+1}\widehat{H}_k$.
We choose the first system (with shift $\sigma_1)$ to be
the seed system, i.e., the system for which we run ``true'' restarted
GMRES. This implies that we have to solve a least squares problem
involving $ \widehat{H}_k-\sigma_1 \widehat{I}_k$ to get the corresponding iterate. Here the matrix $\widehat{I}_k$ denotes the $k$-dimensional identity matrix extended with an extra row of zeros.
For the other shifts $\sigma_2,\ldots,\sigma_s$ we impose the
collinearity constraint for the residuals. The corresponding iterates
are now obtained via solutions of linear systems. For a detailed
derivation we refer to \cite{frommer98}, and the detailed algorithmic
formulation is given in Algorithm~\ref{gmres:alg}.

\begin{algorithm}[h]
\begin{alg}\rm Restarted GMRES-LR($m,k$)\label{gmres:alg}
\begin{algorithmic}
\STATE \{{\bf Input} $m$, $k=\kmax$, $A$, $\{\sigma_1, \dots, \sigma_s\}$, $\{\omega_1,\ldots,
\omega_s\}$, $b$, $L=L_m$, $R=R_m$, $\Lambda=\Lambda_m$\}
\STATE
\STATE $x = x_P = R f(\Lambda) {L}^\dagger b$
\STATE $r = (I-P)b$
\STATE $\rho_0^{(i)} = 1$, $i = 2, \dots, s$
\STATE $\beta = \|r\|_2$
\WHILE[\emph{loop over restart cycles}]{not all systems are converged}
\STATE $v_1 = r/\beta$
\STATE compute $V_k$, $\widehat{H}_k$ by running $k$ steps of Arnoldi for
$A$
\STATE compute $y_k^{(1)}$ as the minimizer of $\, \| \beta e_1 - (\widehat{H}_k-\sigma_1 \widehat{I}_k)y \|
_2 $ 
\FOR{$i = 2, \dots, s$}
\STATE compute $y_k^{(i)}$ and $\rho_k^{(i)}$ as the solution of the
$(k+1) \times (k+1)$ system $\, \left[\widehat{H}_k-\sigma_i \widehat{I}_k \left| V_{k+1}^\dagger
r\right.\right] \left[ \begin{array}{c} y_k^{(i)} \\
                         \rho_k^{(i)}
                      \end{array} \right] = \rho_0^{(i)} \beta e_1 $
\ENDFOR
\STATE $x = x + V_k \sum_{i=1}^s \omega_i y_k^{(i)}$
\STATE $r = r - V_{k+1} (\widehat{H}_k - \sigma_1 \widehat{I}_k) y_k^{(1)}$
\STATE $\beta = \|r\|_2$
\STATE $\rho_0^{(i)} = \rho_k^{(i)}$, $i = 2, \dots, s$
\STATE $r = (I-P)r$ \COMMENT{\emph{projection step}}
\ENDWHILE
\end{algorithmic}
\end{alg}
\end{algorithm}

\subsection{Two-sided Lanczos with rational functions}

We now turn to methods based on two-sided Lanczos, corresponding to
2a) in Section~\ref{sec:list}.
In this case there is no need for restarts because the two-sided
Lanczos process uses only short recurrences anyway.  We summarize a
high-level view of the resulting computational method using multishift
BiCG as Algorithm~\ref{bicg:alg}. The changes necessary to obtain
multishift BiCGStab/QMR should be obvious.

\begin{algorithm}[t]
\begin{alg}\rm BiCG-LR($m$)\label{bicg:alg}
\begin{algorithmic}
\STATE \{{\bf Input} $m$, $A$, $\{\sigma_1, \dots, \sigma_s\}$, $\{\omega_1,\ldots,
\omega_s\}$, $b$, $L=L_m$, $R=R_m$, $\Lambda=\Lambda_m$\}
\STATE
\STATE $x_P = R f(\Lambda) {L}^\dagger b$
\STATE $r = (I-P)b$
\FOR{$k=1,2,\ldots$ until all systems are converged}
\STATE compute the $k$-th BiCG iterates $x_k^{(i)}$, $i=1,\ldots,s$, for the
systems $(A-\sigma_i I) x^{(i)} = r$
\ENDFOR
\STATE $x = x_P + \sum_{i=1}^s \omega_i x_k^{(i)}$
\end{algorithmic}
\end{alg}
\end{algorithm}

\subsection{Direct application of the two-sided Lanczos approach}

We now consider the two-sided Lanczos approach for
$f(A)(I-P)b$ as given in \eqref{bloch_approx_unsym:eq}, corresponding
to 2b) in Section~\ref{sec:list}. The resulting
computational method is summarized as
Algorithm~\ref{two-sided-Lanczos-LR:alg}. Note that due to
  the deflation this algorithm uses a modified shadow vector: We
  remove from $b$ all critical eigenvector components belonging to the
  right eigenvectors of $A^\dagger$, i.e., the left eigenvectors of
  $A$. With this modified shadow vector, the biorthogonality relation
  enforced by the two-sided Lanczos process numerically helps keeping
  the computed basis for $K(A,r)$ free of contributions from the right
  critical eigenvectors, as it should be in exact arithmetic. In
  Algorithm~\ref{two-sided-Lanczos-LR:alg}, the parameter $m$ denotes
  the number of deflated eigenvalues, and $k$ is the maximum dimension
  of the Krylov subspace being built, a parameter which has to be
  fixed \emph{a priori}.

\begin{algorithm}[t]
\begin{alg}\rm Direct two-sided Lanczos-LR($m$, $k$)\label{two-sided-Lanczos-LR:alg}
\begin{algorithmic}
\STATE \{{\bf Input} $m$, $k$, $A$, $b$, $L=L_m$, $R=R_m$, $\Lambda=\Lambda_m$\}
\STATE
\STATE $x_P = R f(\Lambda) {L}^\dagger b$
\STATE $r = b - R {L}^\dagger b$
\STATE $\tilde{r} = b - L {R}^\dagger b$
\COMMENT{\emph{the modified shadow vector}}
\STATE put $v_1 = r$, $\beta = \|r\|_2$, choose $w_1 = \tilde{r}$ and
normalize s.t.\ $v_1^\dagger w_1 = 1$
\FOR{$j=1,2,\ldots, k$}
\STATE update $H_j$, compute $v_{j+1}$ and $w_{j+1} $ from the two-sided
Lanczos process \eqref{2sided1:eq},
\eqref{2sided2:eq}
\ENDFOR
\STATE put $x_k = x_P + \beta \cdot V_k f(H_k) e_1$

\end{algorithmic}
\end{alg}
\end{algorithm} 

We now analyze the main contributions to the computational cost of
Algorithm \ref{two-sided-Lanczos-LR:alg}, which will also be used in
our numerical tests.

\begin{prop}\label{prop:2} Let $M_n$ denote the cost for one matrix-vector
  multiplication by the matrix $A$, and let $k_{\rm tot}$ be the total
  number of iterations performed, i.e., $k_{\rm tot}= k$ from
  Algorithm~\ref{two-sided-Lanczos-LR:alg}. The computational cost of
  Algorithm~\ref{two-sided-Lanczos-LR:alg} is given as
  \begin{equation} \label{cost_2sL:eq}
    2 k_{\rm tot}M_n + \mathcal{O}(k_{\rm tot}n) + \mathcal{O}(mn) + 
    \mathcal{O}(k_{\rm tot}^3).
  \end{equation}     
\end{prop}
To see this, we discuss the dominating contributions to the
computational cost as we did for Algorithm~\ref{fom:alg}.  The
initialization phase has cost $\mathcal{O}(mn)$, since $R, L \in
\CC^{n\times m}$. In each sweep through the for-loop, updating $H_j$
and the Lanczos vectors has cost $2 M_n + \mathcal{O}(n)$, which gives
a total of $2 k_{\rm tot} M_n + \mathcal{O}(k_{\rm tot}n)$.  The last
line of the algorithm requires $\mathcal{O}(k_{\rm tot}^3)$ operations
to compute $f(H_{k_{\rm tot}})$ and additional $\mathcal{O}(k_{\rm
  tot}n)$ operations to get $x_{k_{\rm tot}}$.

\section{Choice of the rational function}
\label{sec:rat}

\begin{figure}[t]
  \includegraphics[width=0.47\textwidth]{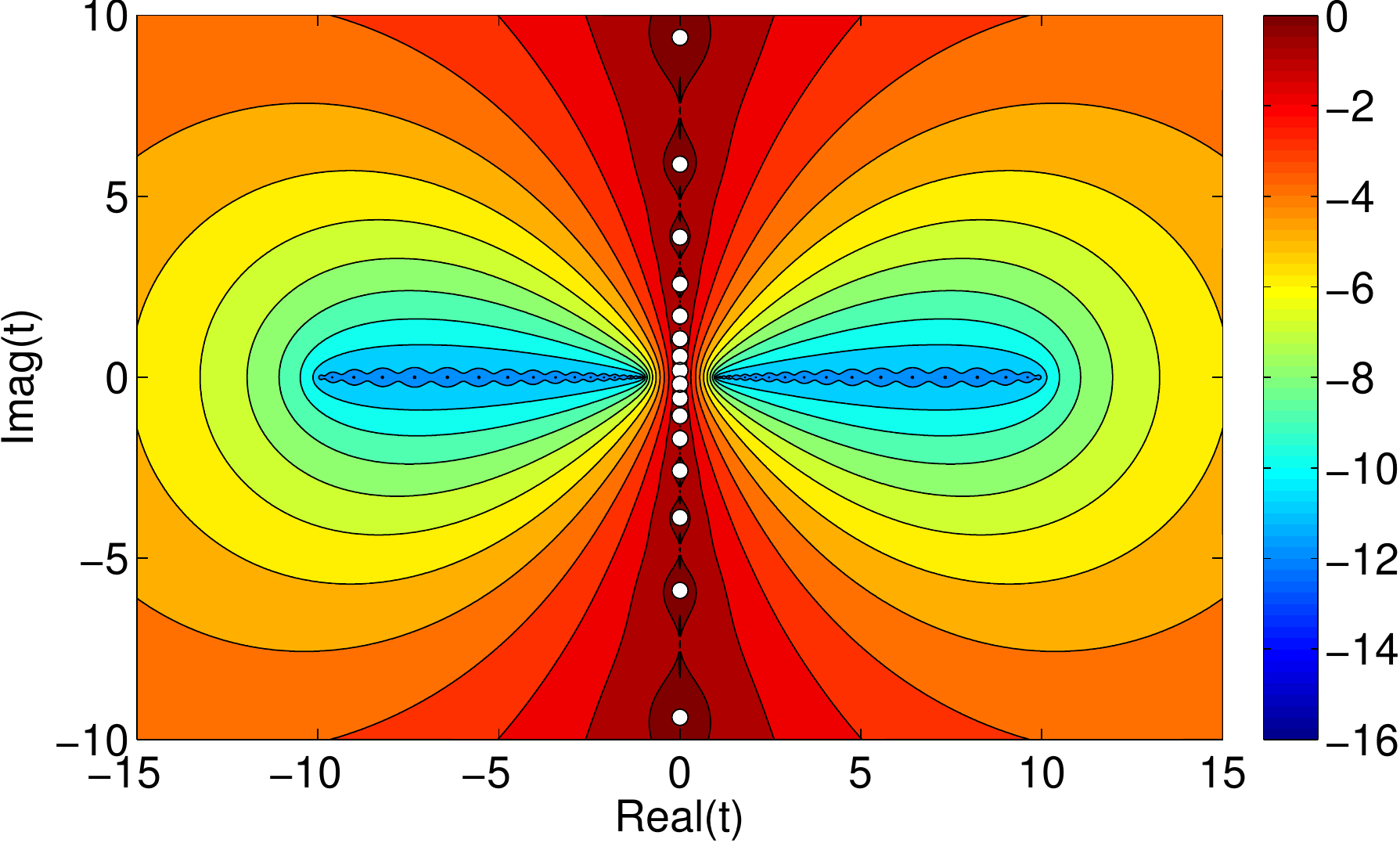}\hfill
  \includegraphics[width=0.47\textwidth]{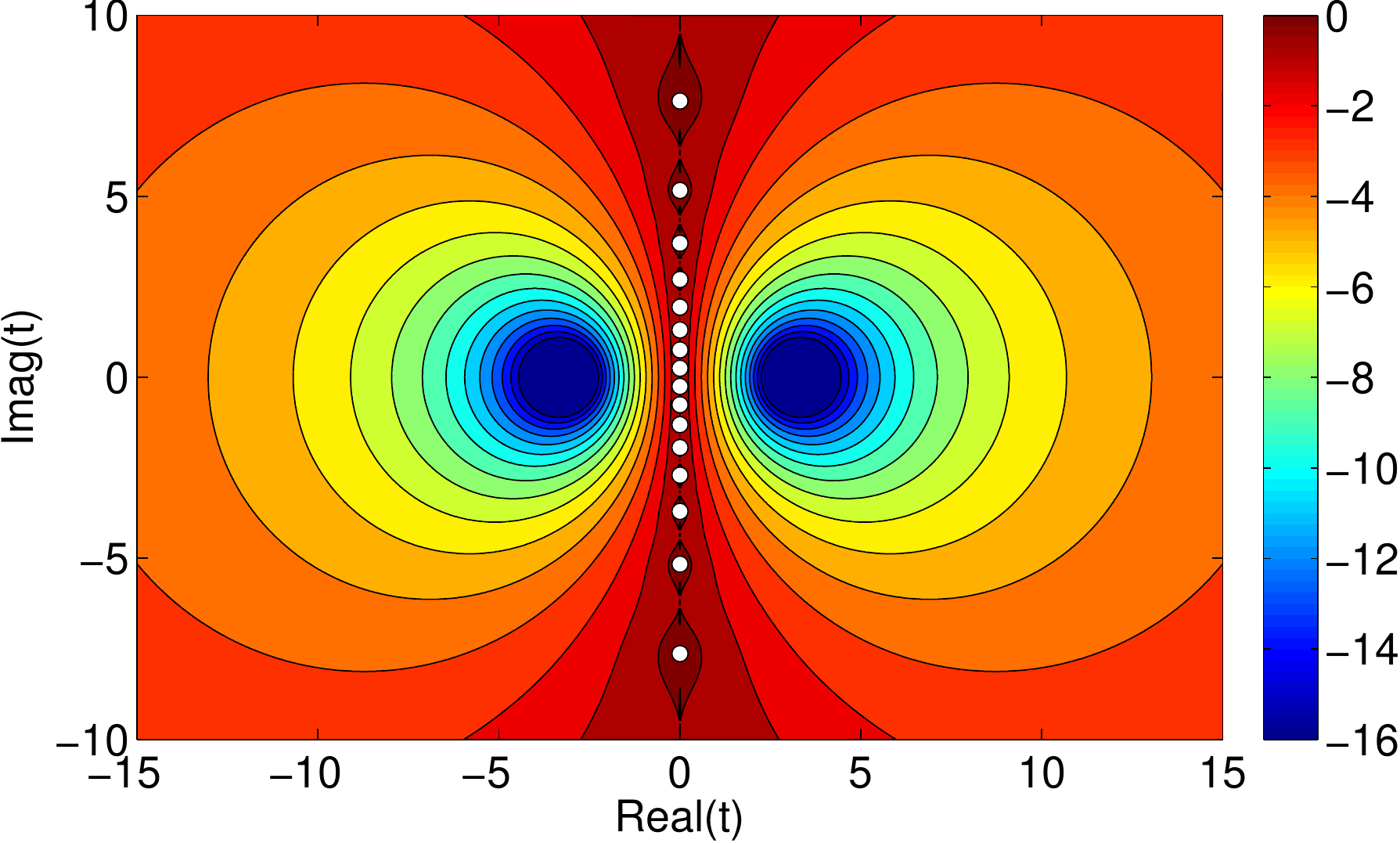}
  \caption{\label{error:fig}Error of the Zolotarev rational
    approximation (left) and the Neuberger rational approximation
    (right). Both rational approximations are of the form $g_s(t) = t
    \sum_{i=1}^s \omega_i/(t^2 - \sigma_i)$ with $\sigma_i < 0$. We
    took $s=10$ in both cases and plotted the contours for
    $\log_{10}(g_{10}(t) - \sign(t))$. We chose $a = 1$ and $b = 10$
    for the Zolotarev approximation, and $c = 1/\sqrt{10}$ for the
    Neuberger approximation.  The white spots on the imaginary axis
    mark the poles $t=\pm i \sqrt{-\sigma_i}$ of the rational
    approximation that lie in the interval $i[-10,10]$.}
\end{figure}

In this section we address the issue of how to find good rational
approximations to the sign function in the non-Hermitian case.

In the Hermitian case, if we know intervals $[-b,-a]$, $[a,b]$ which
contain the (deflated) spectrum of $A$, the sign function of $A$ can
be approximated using the Zolotarev best rational approximation, see
\cite{zolotarev77} and, e.g., \cite{ingerman00,
  vandenEshof:2002ms}. Using the Zolotarev approximation on
non-Hermitian matrices gives rather poor results, unless all
eigenvalues are close to the real axis (see the left plot in
Figure~\ref{error:fig}).  A better choice for generic non-Hermitian
matrices is the rational approximation originally suggested by Kenney
and Laub \cite{KL} and used by Neuberger \cite{Neuberger:1998my,
  Neuberger:1999zk} for vanishing chemical potential,
\begin{equation} \label{neuberger_mit_c:eq}
  \sign(t) \approx g_s(ct)\,, \mbox{ where }  
   g_s(t) = \frac{(t+1)^{2s} - (t-1)^{2s}}{(t+1)^{2s} 
    + (t-1)^{2s}}
\,.
\end{equation}
Note that $g_s(t)=g_s(1/t)$, and $g_s(t) = \tanh\left(2s \; \text{atanh}\, t\right)$ for $|t|<1$.
The partial fraction expansion of $g_s$ is known to be
\begin{equation} \label{pfe_s:eq}
  g_s(t) = t \sum_{i=1}^s \frac{\omega_i}{t^2 - \sigma_i} \quad
  \text{with } 
  \omega_i = \frac{1}{s} \cos^{-2}\left( \frac{\pi}{2 s}
    \left(i-\frac{1}{2}\right)\right)\,, \quad \sigma_i =
  -\tan^2\left(\frac{\pi}{2s}\left(i-\frac{1}{2}\right)\right)\,, 
\end{equation}
see \cite{KL,Neuberger:1998my}.  In \eqref{neuberger_mit_c:eq}, $c>0$
is a parameter which one chooses to minimize the number of poles $s$ of
the partial fraction expansion \eqref{pfe_s:eq}, 
see \cite{vandenEshof:2002ms} and the discussion after Theorem~\ref{circle:thm} below.
Whereas for the Zolotarev approximation the regions of good
approximation are concentrated along the real axis, the approximation
$g_s(t)$ approaches $\sign(t)$ well on circles to the left and right of
the imaginary axis, see the right plot in Figure~\ref{error:fig}.
For this reason, the Neuberger approximation is better suited for
generic non-Hermitian matrices.  All we need is some {\em a priori}
information on the spectrum from which we can determine an appropriate
circle $C$ in the right half-plane, centered on the real axis, such
that $C$ together with $-C$ contains all the eigenvalues.  We then can
compute the degree $s$ of the Neuberger approximation such that the
sign function is approximated to a given accuracy on $C \cup -C$.

The following theorem gives the insight necessary for this approach to work.
\begin{theorem} \label{circle:thm}
For given $s$ and $\epsilon>0$ we have 
\begin{equation}
  |e_s(t)| = |g_s(t) - \sign(t)| \le \epsilon \quad \text{for } t \in
     C_{s,\epsilon} \text{ or } -t \in C_{s,\epsilon}\,, 
 \label{eq:eps}
\end{equation}
where $C_{s,\epsilon}$ is the circle with radius $R = 2
\delta(\epsilon, s)/[\delta(\epsilon,s)^2 - 1]$ and center $M =
[\delta(\epsilon,s)^2 + 1]/[\delta(\epsilon,s)^2 - 1]$,
with $\delta(\epsilon,s) = (2/\epsilon + 1)^{1/(2s)}$.
\end{theorem}
\emph{Proof.}  Assume that $t$ is in the right half-plane
  (the case of $t$ in the left half-plane can be treated in a
  completely analogous manner). With $z = [(t+1)/(t-1)]^{2s}$ we write
$g_s(t) = (z - 1)/(z + 1)$ such that $e_s(t) = g_s(t) - 1 = -2/(z +
1)$. Therefore $|e_s(t)| \le \epsilon$ if and only if $|z + 1| \ge
2/\epsilon$.

Since $|z| - 1 \le |z + 1|$, a sufficient condition for $|e_s(t)| \le
\epsilon$ is $|z| - 1 \ge 2/\epsilon$, which is equivalent to
\begin{equation}
  \left| \frac{t+1}{t-1} \right| \ge \left(\frac{2}{\epsilon} +
    1\right)^{1/(2s)} = \delta(\epsilon,s)\,. 
\end{equation}
Let $t = x + i y$ be on the circle $C_{s,\epsilon}$, i.e., $(x - M)^2
+ y^2 = R^2$. Then
\begin{align}
  \left| \frac{t+1}{t-1} \right|^2 &= \frac{(x+1)^2 + y^2}{(x-1)^2 +
    y^2}\notag\\
  &= \frac{(x+1)^2 + R^2 - (x - M)^2}{(x-1)^2 + R^2 - (x -
    M)^2}\notag\\
  &= \frac{2x(M+1) + 1 + R^2 - M^2}{2x(M-1) + 1 + R^2 - M^2}\,.
\end{align}
In fact we have $1 + R^2 - M^2 = 1 + (R-M)(R+M) = 1 -
\frac{\delta(\epsilon,s)-1}{\delta(\epsilon,s)+1} \cdot
\frac{\delta(\epsilon,s)+1}{\delta(\epsilon,s)-1}=0$ and thus
\begin{equation}
  \left| \frac{1 + t}{1 - t} \right|^2 = \frac{M+1}{M-1}
  = \frac{\frac{\delta(\epsilon,s)^2+1}{\delta(\epsilon,s)^2-1}+1}
  {\frac{\delta(\epsilon,s)^2+1}{\delta(\epsilon,s)^2-1}-1} 
  = \delta(\epsilon,s)^2\,.
\end{equation}
So we have shown that $|e_s(t)| \le \epsilon$ on the boundary of the
circle $C_{s,\epsilon}$, and by the maximum modulus principle this also holds
for $t$ inside the circle. \hfill $\Box$ \smallskip

The parameter $c$ in (\ref{neuberger_mit_c:eq}) can now be used 
in order to optimize the number of poles for a given target
accuracy $\epsilon$ if the
spectrum of the operator is known to be contained in the union of two circles $C(m,r) \cup C(-m,r)$,
where $C(m,r)$ is the circle $\{ |z-m| \leq r\}$ and $m$ is real, 
$0 < r <m$.  
For symmetry reasons it is again sufficient to discuss only the circle
in the right half-plane, $C(m,r)$. 
Note that $s$ is a positive integer.  Restricting the function $g_s(t)$ 
to real arguments, we see that it is positive on $(0,\infty)$, 
monotonically increasing on $t \in (0,1]$, and that
$g_s(t) = g_s(1/t)$ as well as $g_s(1) = 1$. 
The maximum error $e_{\max} = \max_{t \in [m-r,m+r]} |1-g_s(ct)|$ is therefore smallest if $c$ is chosen such that the scaled interval $[c(m-r),c(m+r)]$ 
is of the form $[1/d,d]$. This is the case for 
$c = ((m+r)(m-r))^{-1/2}$ with $d = ((m+r)/(m-r))^{1/2}$, see also \cite{vandenEshof:2002ms}.\footnote{We thank an anonymous referee for pointing out that this discussion also shows that we could reduce the error from $e_{\max}$ to $e_{\max}/(2-e_{\max})$ if we multiplied $g_s(t)$ by $\alpha=2/(2-e_{\max})$.}  
For
this choice of $c$ we see that $t$ is in $C(m,r)$ if and only if 
$ct$ is in $C(M,R)$ with $M=\frac{d^2+1}{2d}$ and $R=\frac{d^2-1}{2d}$.
But $C(M,R)$ is precisely of the form that was considered in Theorem~\ref{circle:thm} with $\delta(\epsilon,s) = \frac{d+1}{d-1}$.
Therefore, if we want the error $|g_s(ct)-1|$ to be smaller than $\epsilon$
 for $t \in C(m,r)$, Theorem~\ref{circle:thm} tells us that it is 
sufficient to require $\frac{d+1}{d-1} = \delta(\epsilon,s) = (2/\epsilon + 1)^{1/(2s)}$. Solving for $s$ we see that this precision is obtained if the number $s$ of poles satisfies 
\begin{equation}\label{no_of_poles:eq}
s \ge  \frac{\log\left(\frac{\epsilon}{\epsilon + 2}\right)}{2 \cdot \log\left(\frac{d-1}{d+1}\right)} .
\end{equation}

\section{Numerical results} \label{example:sec}

This section contains the results of several numerical experiments
comparing some of the methods developed in this paper. We only present
results for Algorithms~\ref{fom:alg} and
\ref{two-sided-Lanczos-LR:alg} since the results for
Algorithms~\ref{gmres:alg} and \ref{bicg:alg} are very similar to
those of Algorithm~\ref{fom:alg} \cite{Schaefer:2008phd}.
Algorithms~\ref{fom:alg} and \ref{two-sided-Lanczos-LR:alg} as
described in Section~\ref{sec:alg} were applied to compute
$\sign(H_W)b$, where $H_W=\gamma_5D_W(\mu)$ is the ``Hermitian''
Wilson Dirac operator at nonzero chemical potential and $b= \left(1,
  \dots, 1\right)$ for generic QCD gauge field configurations on
lattices with sizes $4^4, 6^4, 8^4$, and $10^4$.  The lattice
parameters are $\beta=5.1$, $m_W=-2$, $m_q=0$, and $\mu=0.3$, see
\cite{Bloch:2006cd} for the notation.

In Algorithm~\ref{fom:alg} one has to decide which rational
approximation to use.  This decision should be made depending on the
spectrum of $A$. Even though in lattice QCD the eigenvalues do not
move far away from the real axis for reasonable values of $\mu$, we
adopted a conservative strategy and used the Neuberger approximation
in our numerical experiments.  As we discussed at the end of
section~\ref{sec:rat}, in order to use a Neuberger rational
approximation we have to determine circles $C(m,r)$ and $C(-m,r)$
which should contain all the eigenvalues (except the ones that have
been deflated). Of course, we cannot precompute the whole spectrum, so
we have to rely on a reasonable heuristics. From the deflation process
we know a parameter $\alpha >0$ such that all non-deflated eigenvalues
have modulus larger than $\alpha$.  We also precomputed the eigenvalue
which is largest in modulus with value $\beta > 0$. The heuristics,
which is confirmed by additional numerical experiments, is to assume
that for reasonable values of $\mu$ all eigenvalues are contained in
the two circles centered on the real line and intersecting it at the
points $\alpha$, $\beta$ and $-\alpha$, $-\beta$, respectively.  This
gives $m = (\alpha+\beta)/2 $ and $r=(\beta-\alpha)/2$. The number of
poles to use is now given by (\ref{no_of_poles:eq}) together with the
corresponding (scaled) Neuberger approximation.  Note that this
approach is quite defensive since it allows eigenvalues to deviate
substantially from the real axis if their real parts are not close to
$\alpha$, $\beta$, $-\alpha$ or $-\beta$.  For larger lattice volumes
and $\mu$ relatively small we observed that the Zolotarev
approximation based on the intervals $[-\beta,-\alpha]$ and
$[\alpha,\beta]$ can be an interesting alternative, since the spectrum
deviates only marginally from the real axis.  Using Zolotarev instead
of Neuberger would reduce the computational cost for the restarted FOM
method since the number of poles $s$ would be reduced (since this
moves the smallest shifts away from the origin, it also leads to a
reduction in $k_\text{tot}$).  However, as mentioned above, we only
used the more conservative Neuberger approximation.

In Algorithm~\ref{fom:alg} one also has to decide when the iteration
to solve any of the linear systems is considered to be converged.  We
require the norms of the residuals to be less than $\epsilon$, with
$\epsilon$ the target accuracy of the rational approximation defined
in Eq.~\eqref{eq:eps}.  This gives an upper bound of
$\approx2\epsilon$ on the total error.  In our experiments we observed
that the total error (as defined in the next paragraph) was smaller
(as small as $0.1\epsilon$), which is natural since most of the
eigenvalues are in the interior of the circles $C(m,r)$ and $C(-m,r)$,
where the approximation works better than at the boundary.

\begin{figure}[p]
  \centerline{
    \begin{tabular}{c@{\hspace*{15mm}}c}
      \includegraphics[width=0.4\textwidth]{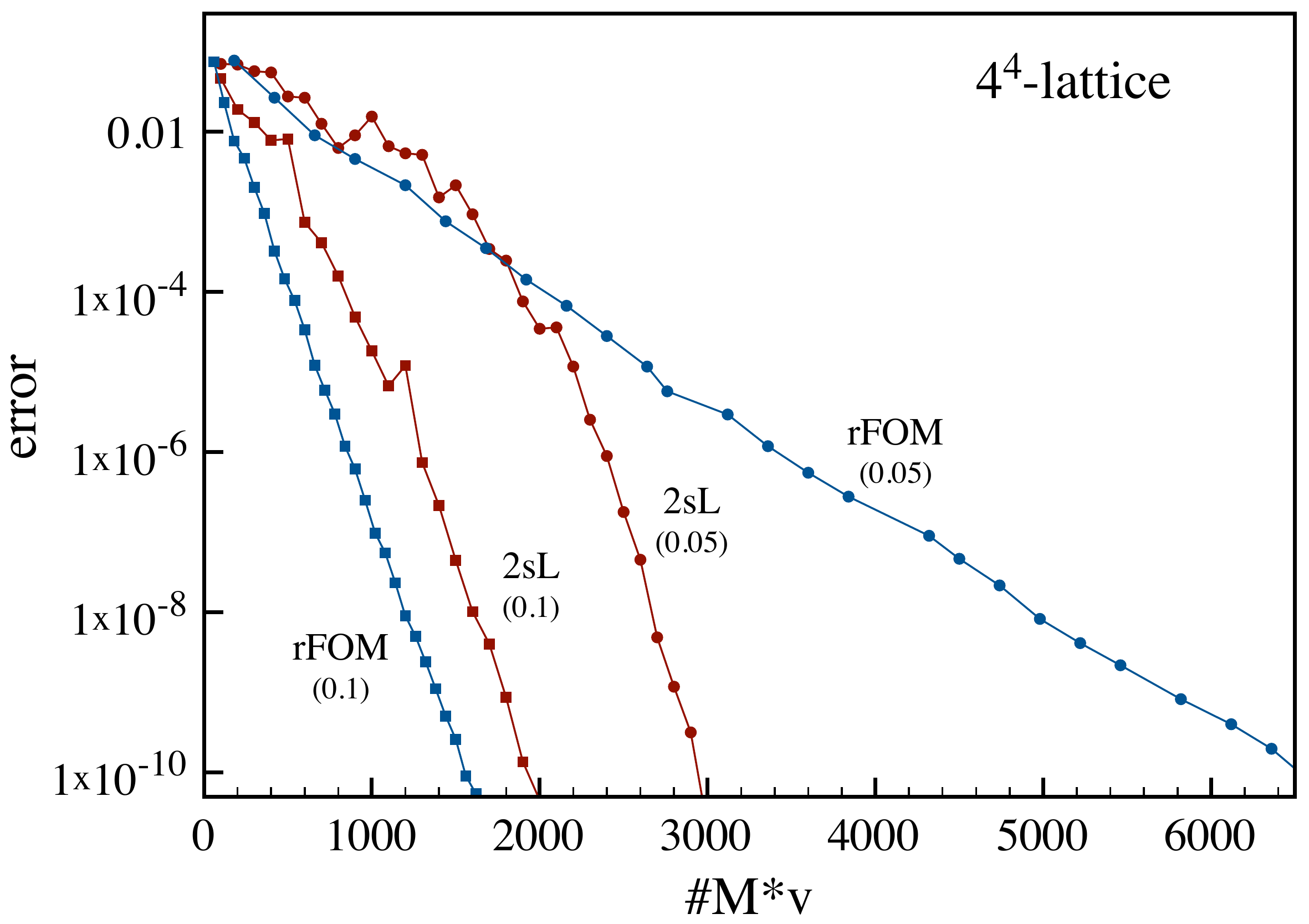} &
      \includegraphics[width=0.4\textwidth]{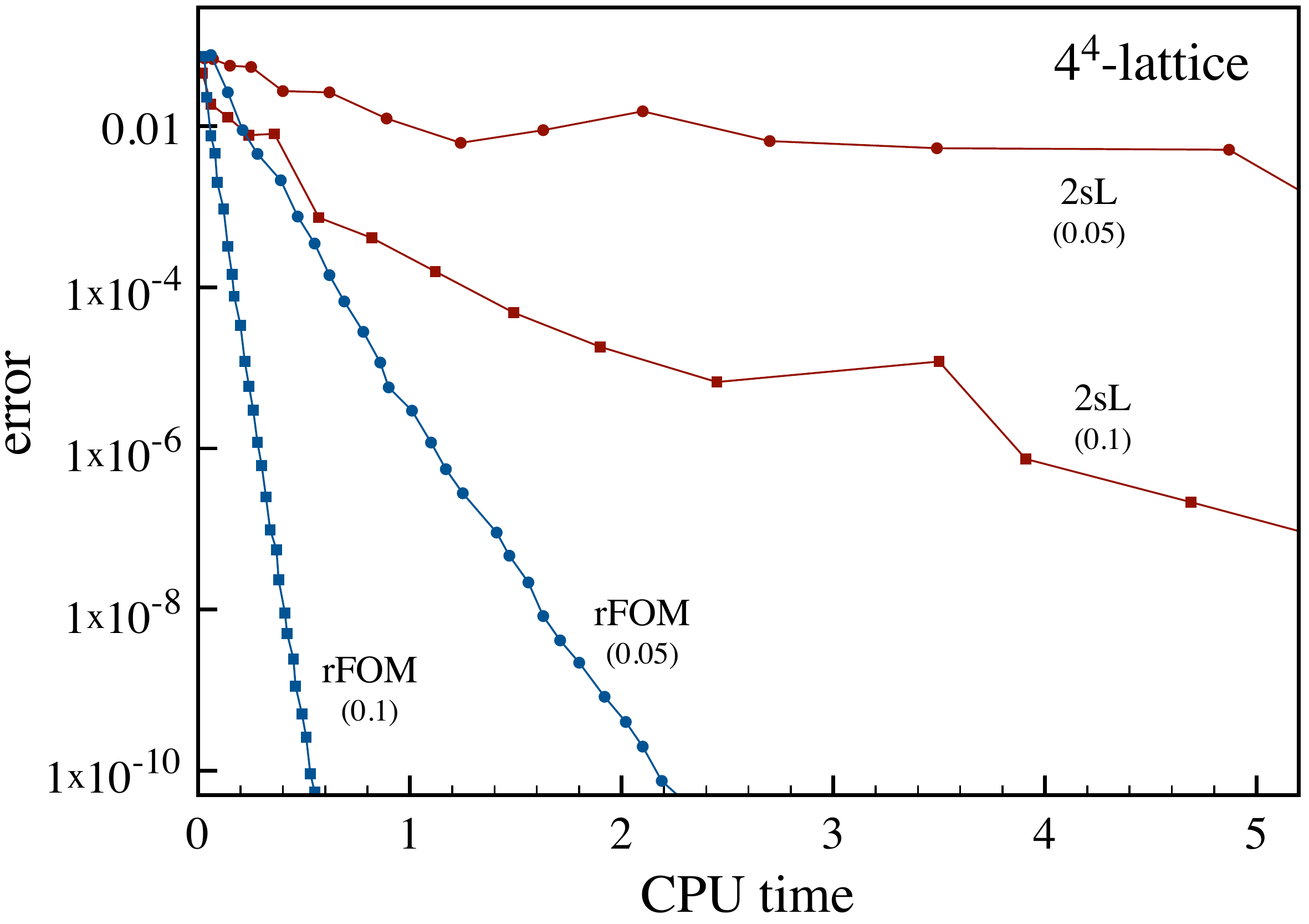}\\[2mm]
      \includegraphics[width=0.4\textwidth]{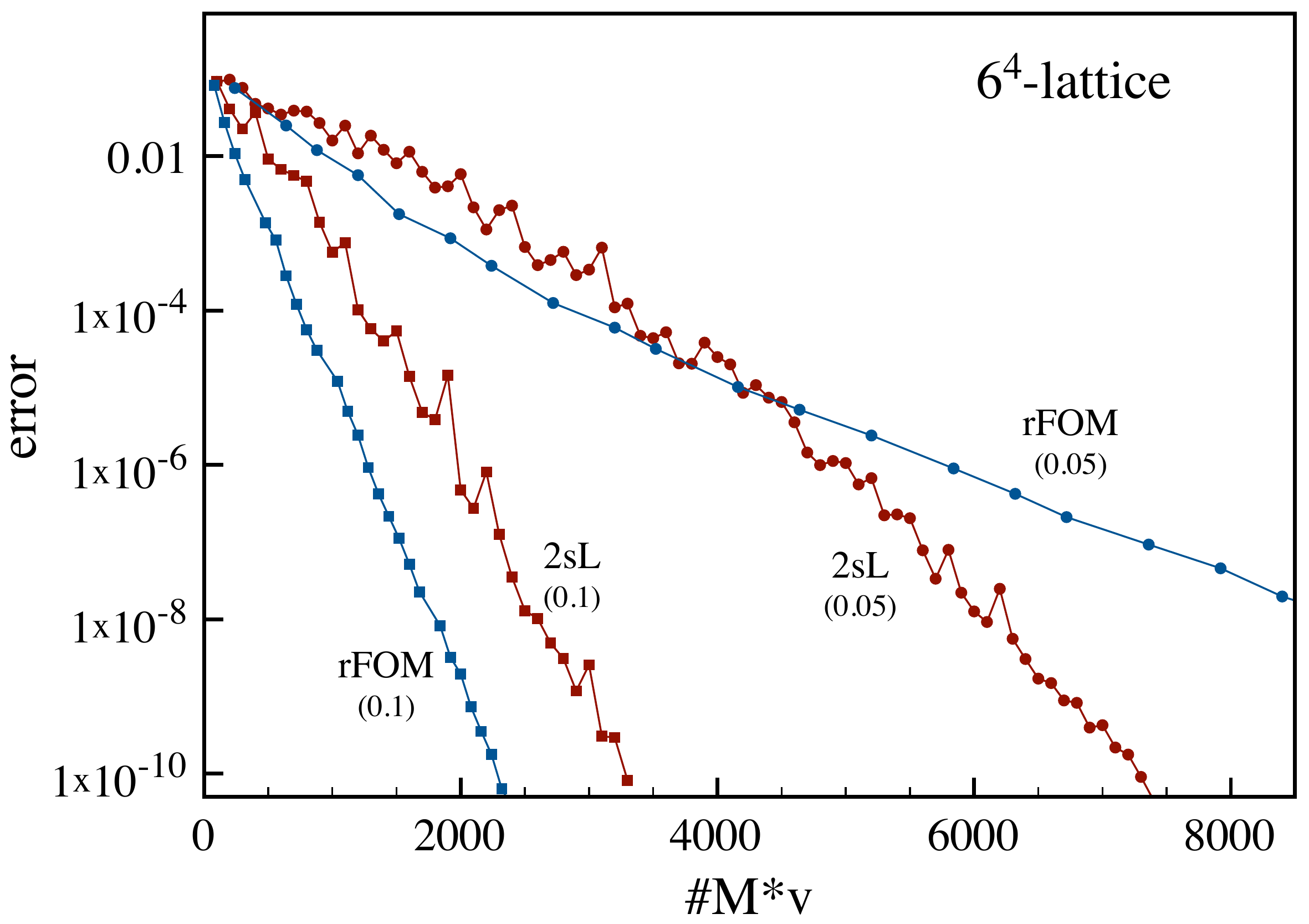} &
      \includegraphics[width=0.4\textwidth]{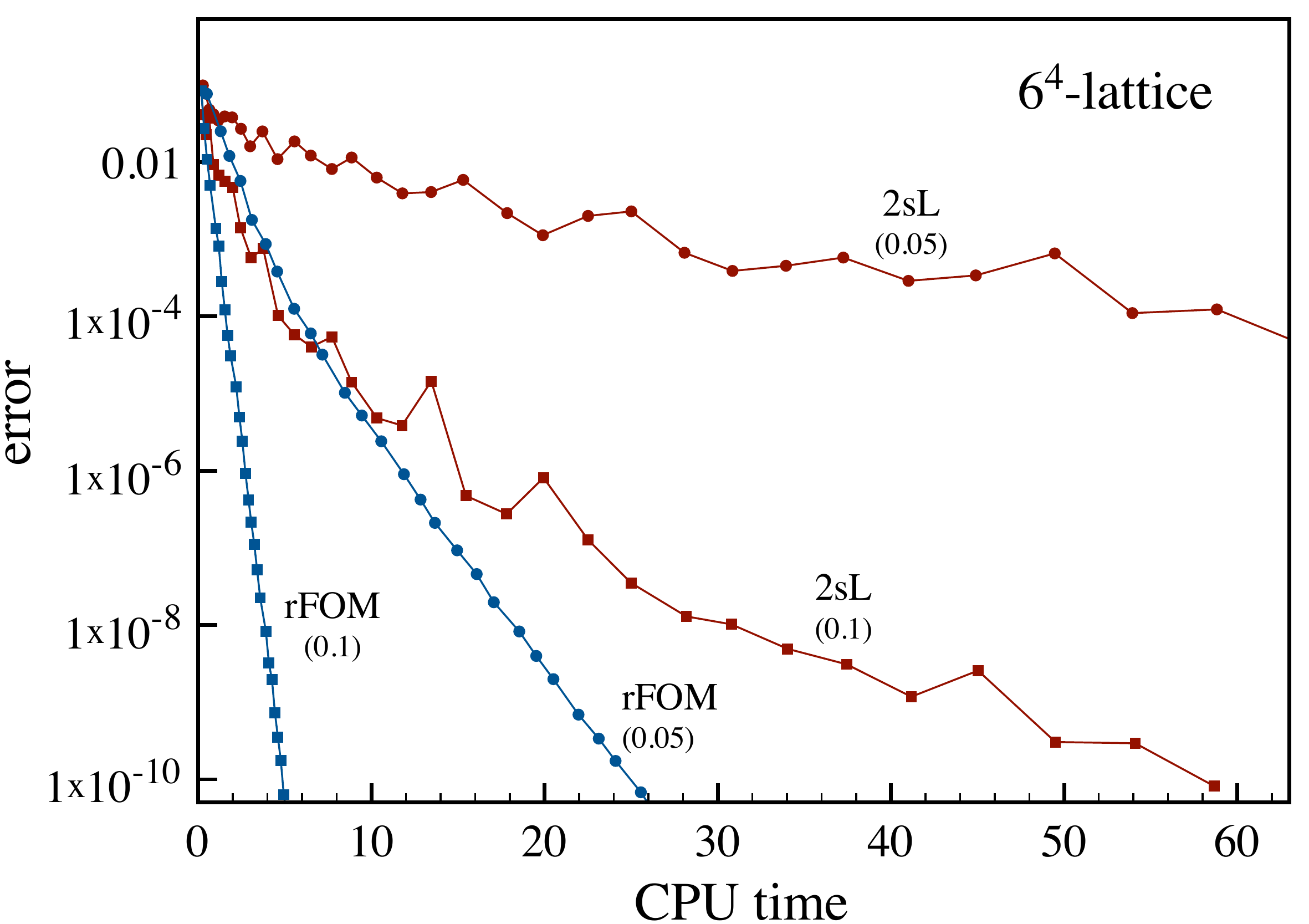}\\[2mm]
      \includegraphics[width=0.4\textwidth]{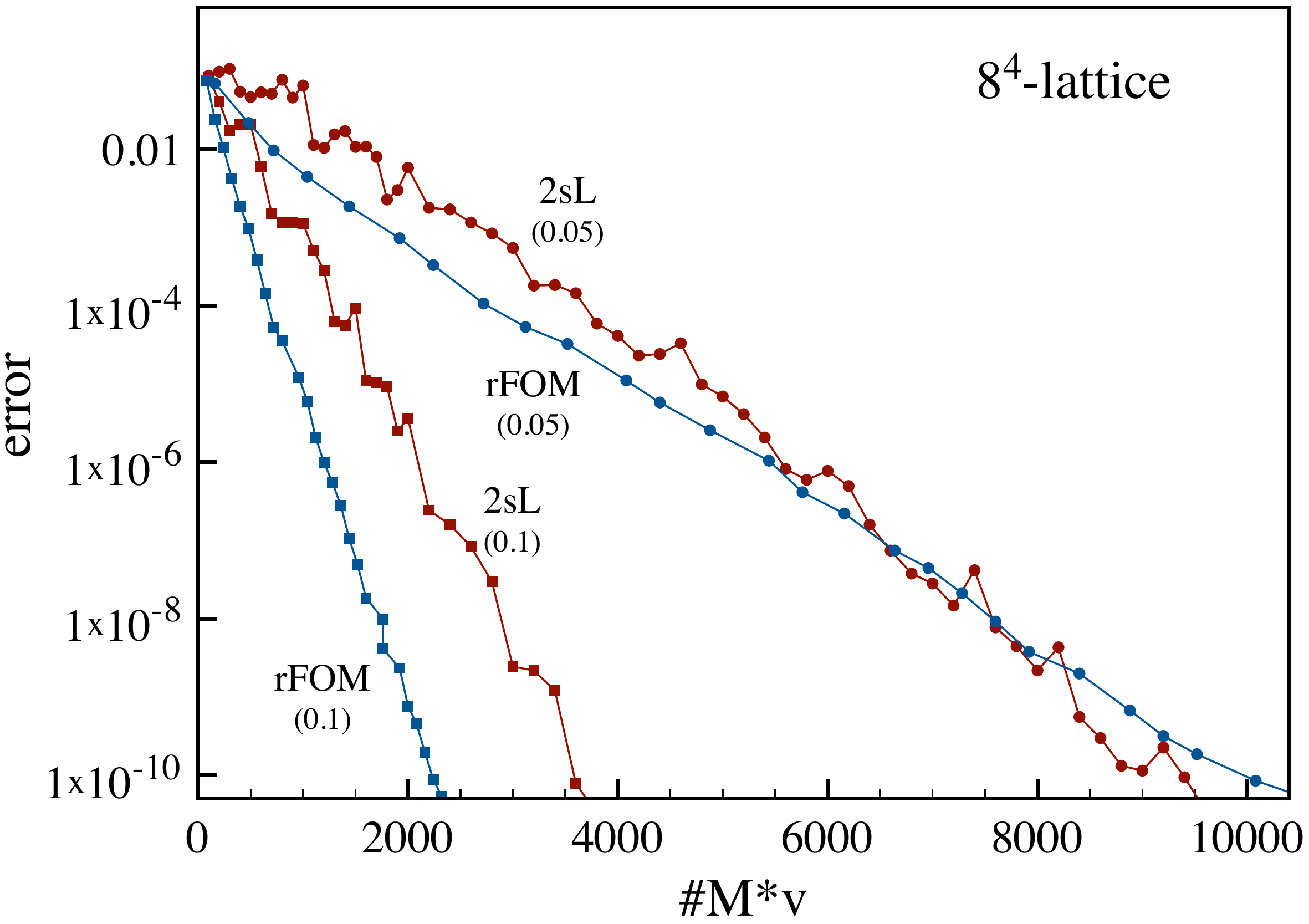} &
      \includegraphics[width=0.4\textwidth]{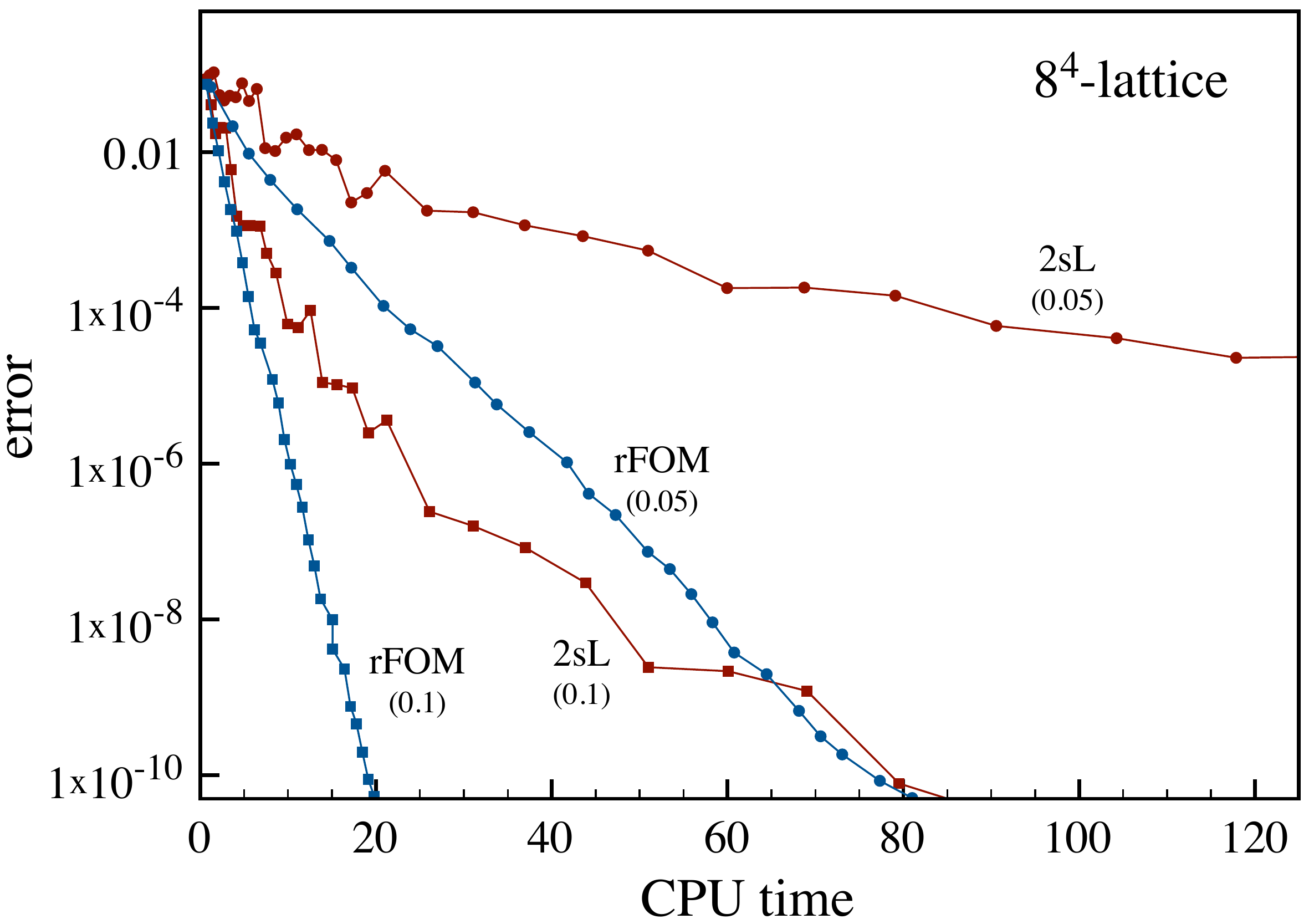}\\[2mm]
      \includegraphics[width=0.4\textwidth]{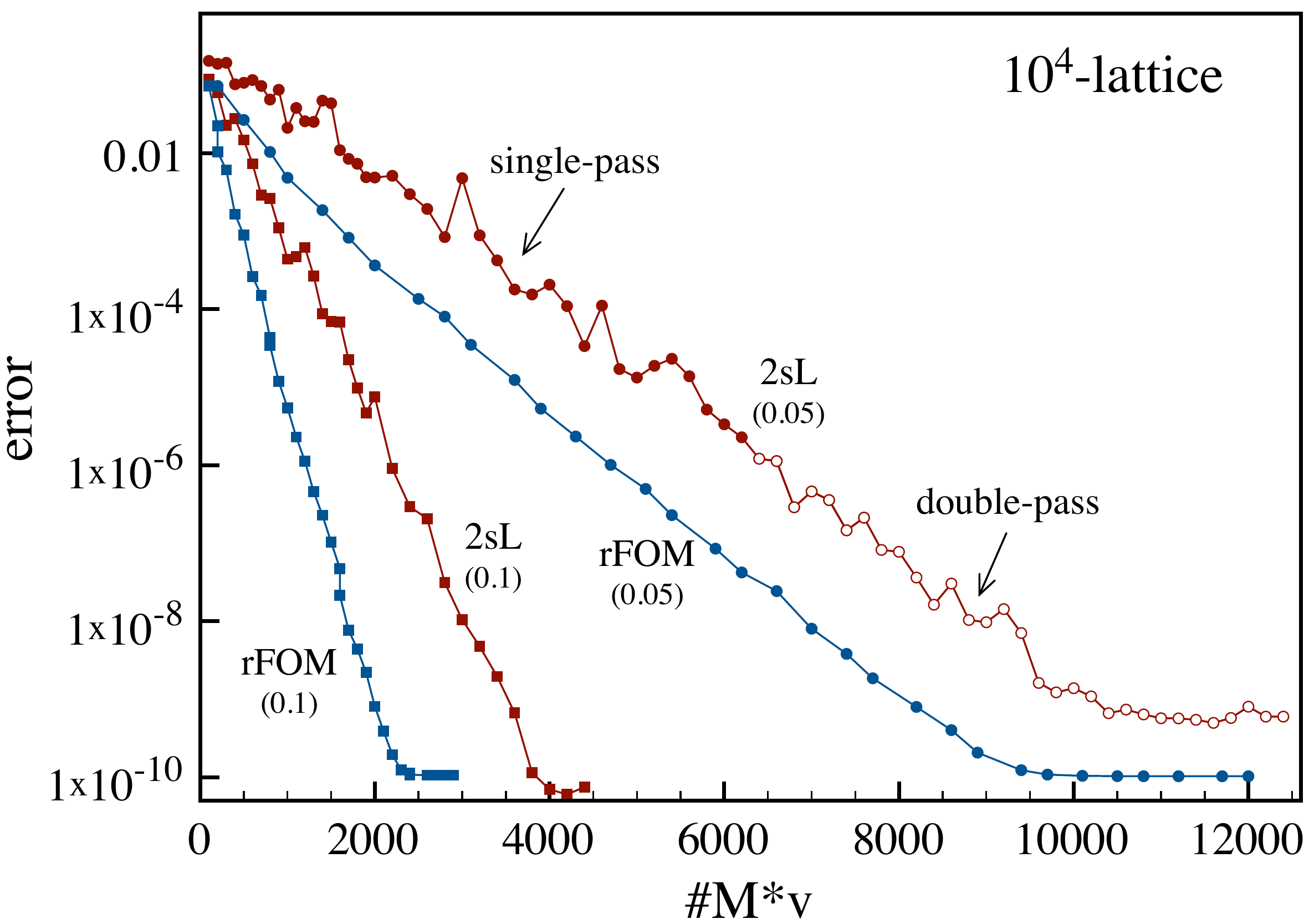} &
      \includegraphics[width=0.4\textwidth]{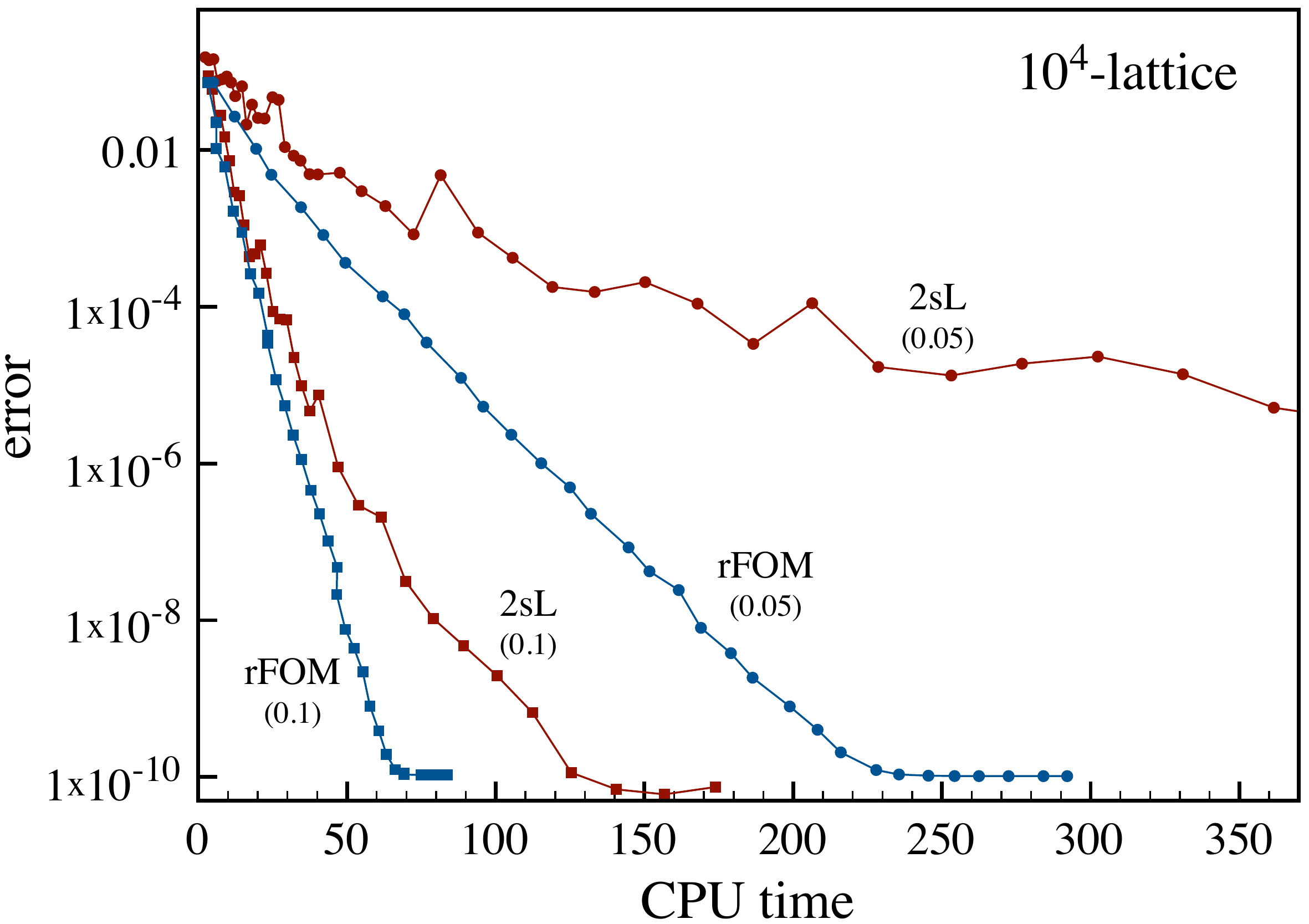}
    \end{tabular}}
  \caption{\label{Fig:acc_vs_cpu}Comparison of the accuracy of the
    restarted FOM-LR algorithm (rFOM) and the direct two-sided
    Lanczos-LR method (2sL) as
    a function of the number of matrix-vector multiplications (left)
    and the CPU time in seconds (right) for a $4^4$ (row 1), $6^4$
    (row 2), $8^4$ (row 3), and $10^4$ (row 4) lattice configuration.
    Each plot shows data for two different deflation gaps, given in parentheses.  The restart size used in the
    restarted FOM-LR algorithm is $\kmax=30$ for the $4^4$ lattice, $\kmax=40$ for the $6^4$ and
    $8^4$ lattices and $\kmax=50$ for the $10^4$ lattice.}
\end{figure}

We now turn to the question of how to determine the accuracy of the
approximations to the sign function in our numerical tests. The exact
error cannot be determined because the computational cost to evaluate
$\sign(A)b$ exactly by a direct method is too large if $A$ is
large. To obtain an estimate for the error, we compute $\sign(A)^2b$
(by applying $\sign(A)$ twice in succession), which should equal $b$
if the approximation to the sign function were exact, and then take
$\frac12||\sign(A)^2b-b||/||b||$ as a measure for the error (or accuracy).
Of course, in production runs one would check the quality of the
approximation only occasionally.

In Figure~\ref{Fig:acc_vs_cpu} we compare the results of the restarted
FOM-LR approximation with those of the direct two-sided Lanczos-LR
method for various lattice sizes: $4^4$ ($n = 3,072$), $6^4$ ($n =
15,552$), $8^4$ ($n = 49,152$), and $10^4$ ($n = 120,000$), and for
two different deflation gaps, i.e., the modulus of the smallest
non-deflated eigenvalue.\footnote{The plots in
  Figure~\ref{Fig:acc_vs_cpu} are for a single configuration per
  volume.  One might ask to what extent this configuration is typical.
  In the present context the main difference between configurations
  lies in the magnitude of their smallest Dirac eigenvalues.  The
  removal of the latter by deflation makes the configuration
  typical.}$^,$%
\footnote{Note that the cost of deflation, i.e., the cost to compute
  the $m$ critical eigenvalues and eigenvectors, is not included in
  these figures (and in the figures below) because it only needs to be
  paid once for each $A$.  In the case of lattice QCD, $\sign(A)b$ has
  to be computed for many different $b$ in an iterative inverter.  One
  should then choose $m$ such that the total run time, including the
  cost of deflation (which strongly depends on the details of $A$), is
  minimized.  However, this optimization issue is not the focus of the
  current paper.}  The accuracy is shown as a function of the number
of matrix-vector multiplications (left) and as a function of the CPU
time on a 2.4 GHz Intel Core 2 with 8 GB of memory (right).  Although
the number of matrix-vector multiplications is often used to compare
the efficiency of different iterative methods, it is not the best
measure of the efficiency since it only includes the first term in
Eqs.~\eqref{cost_fom:eq} and \eqref{cost_2sL:eq},
respectively.\footnote{Note that FOM-LR applied to
  Eq.~\eqref{pfe_s:eq} works with $A^2$, so we actually have
  $C_n=2M_n$.}  The total run time is a better measure since it
includes the other terms as well.  Depending on the parameters
actually used, some of these terms can be dominant or negligible.
E.g., the $\mathcal O(k^3)$ term in the two-sided-Lanczos-LR method
becomes dominant when the Krylov subspace grows.  Another example are
the $\mathcal O(mn)$ terms in both algorithms, which reflect the cost
of using the deflated eigenvectors and which could be neglected in all
cases we considered.  Note that for the two smaller lattices a larger
fraction of the problem fits in cache, which leads to a reduction of
the run time.

\begin{figure}[h]
  \centerline{
    \includegraphics[width=0.4\textwidth]{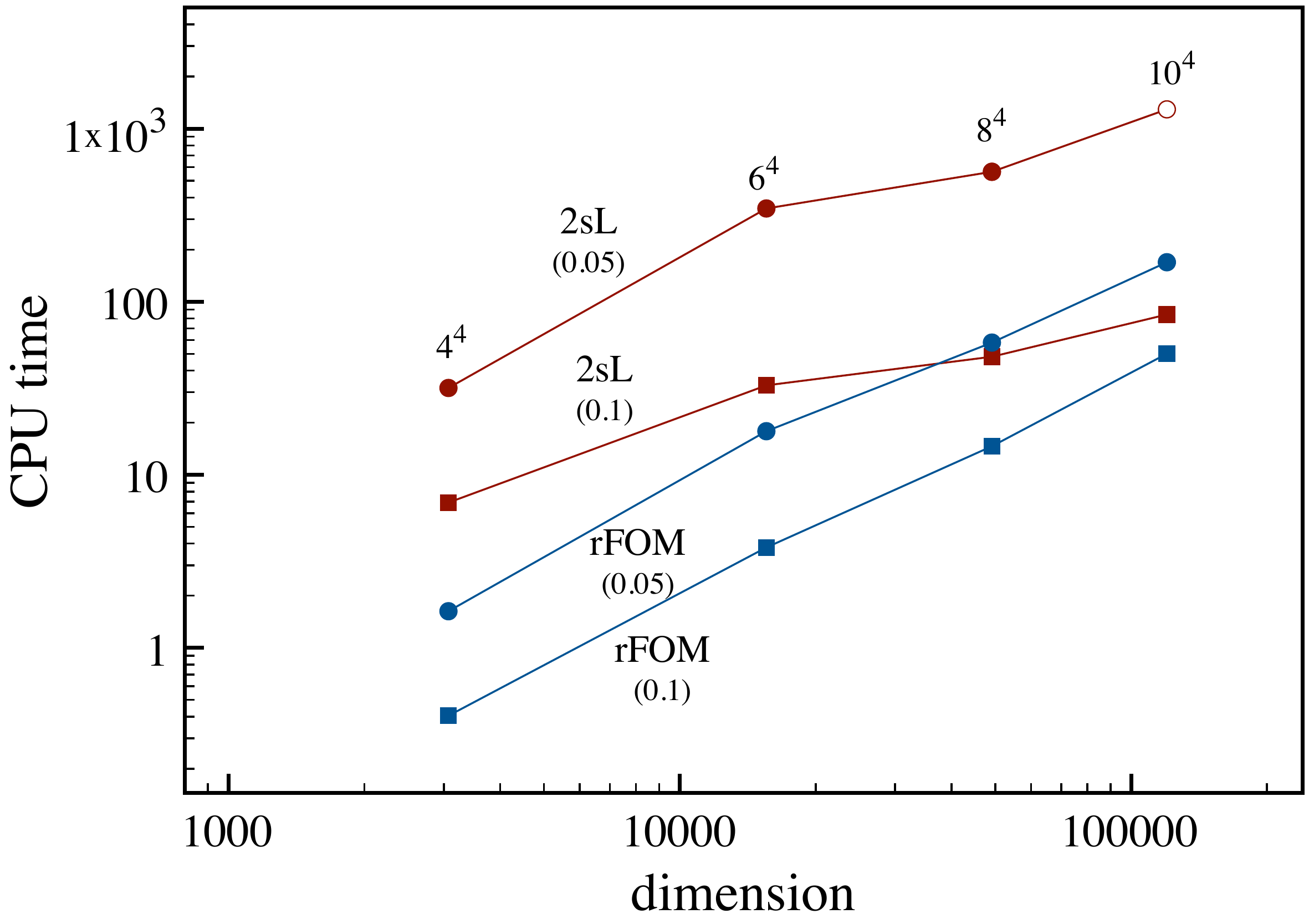}
  }
  \caption{\label{Fig:vol_dep}Run time (in seconds) for the restarted
    FOM-LR algorithm and the direct two-sided Lanczos-LR method as a
    function of the matrix size to achieve an accuracy of $10^{-8}$.
    The run time does not include the cost of deflation.  The
    deflation gaps are given in parentheses, and the restart sizes are
    the same as in Figure~\ref{Fig:acc_vs_cpu}.  The data point for
    2sL(0.05) on a $10^4$ lattice (open circle) was computed in
    double-pass for memory reasons.}
\end{figure}
In Figure~\ref{Fig:vol_dep} we show how the efficiency of both methods
scales with the volume.  This figure should be interpreted with care.
Since we used a constant deflation gap we expect the number of
iterations ($k_\text{tot}$ resp. $k$) to be approximately constant.%
\footnote{This is not necessarily so for the small lattices, where the
  superlinear convergence of Krylov subspace methods might become
  noticeable.} This would result in a contribution to the execution
time which is linearly dependent on the volume, for both
methods. However, there are several effects which obscure this linear
dependence. For example, in the restarted FOM there is a dependence on
$k_{\max}$. In the direct two-sided Lanczos method the ${\cal O}(k^3)$
cost to compute $f(H_k)$ dominates for small volumes.  In addition,
there are the cache effects already mentioned.

\begin{figure}[h]
  \centerline{
    \includegraphics[width=0.4\textwidth]{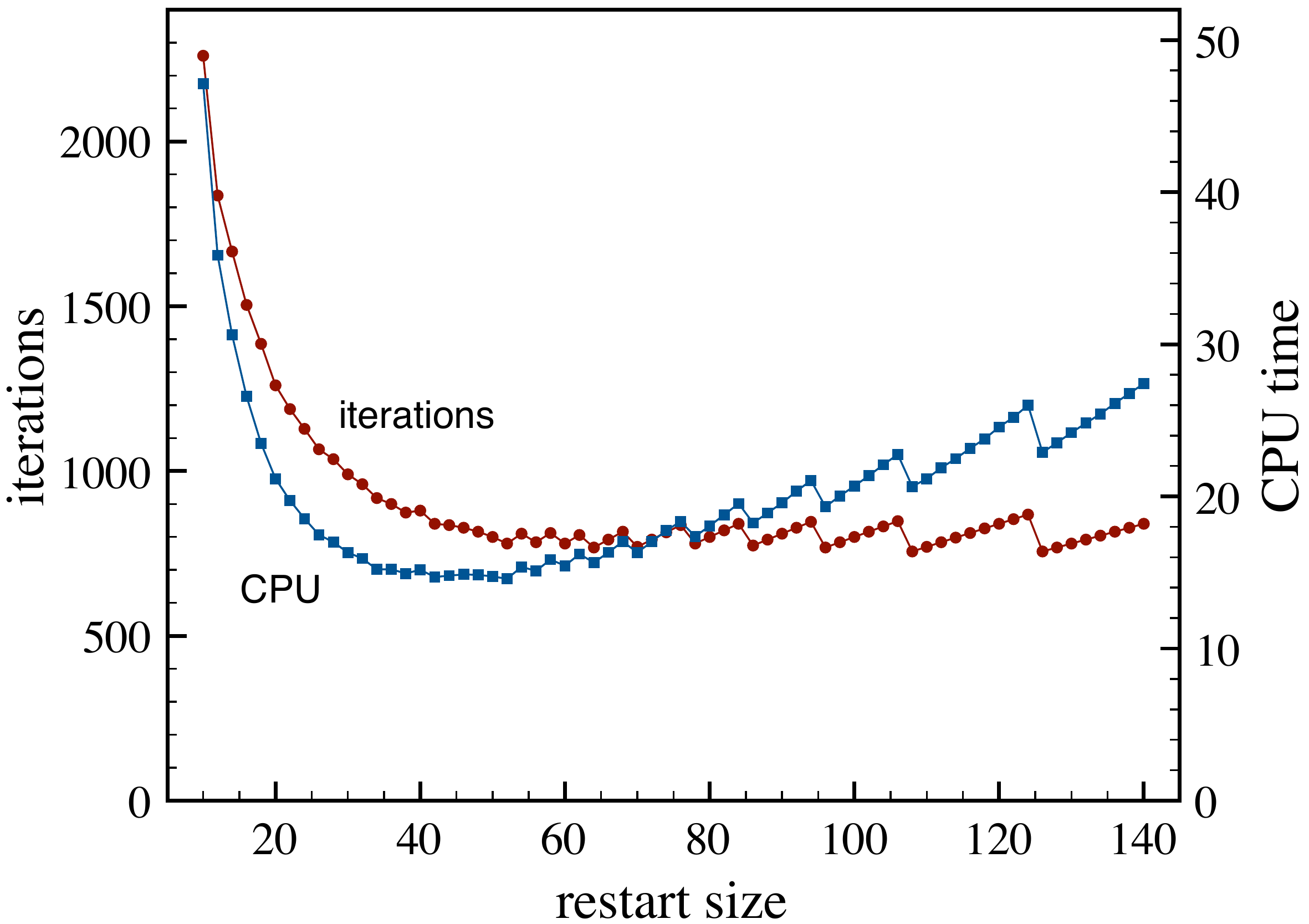}\hspace*{15mm}
    \includegraphics[width=0.4\textwidth]{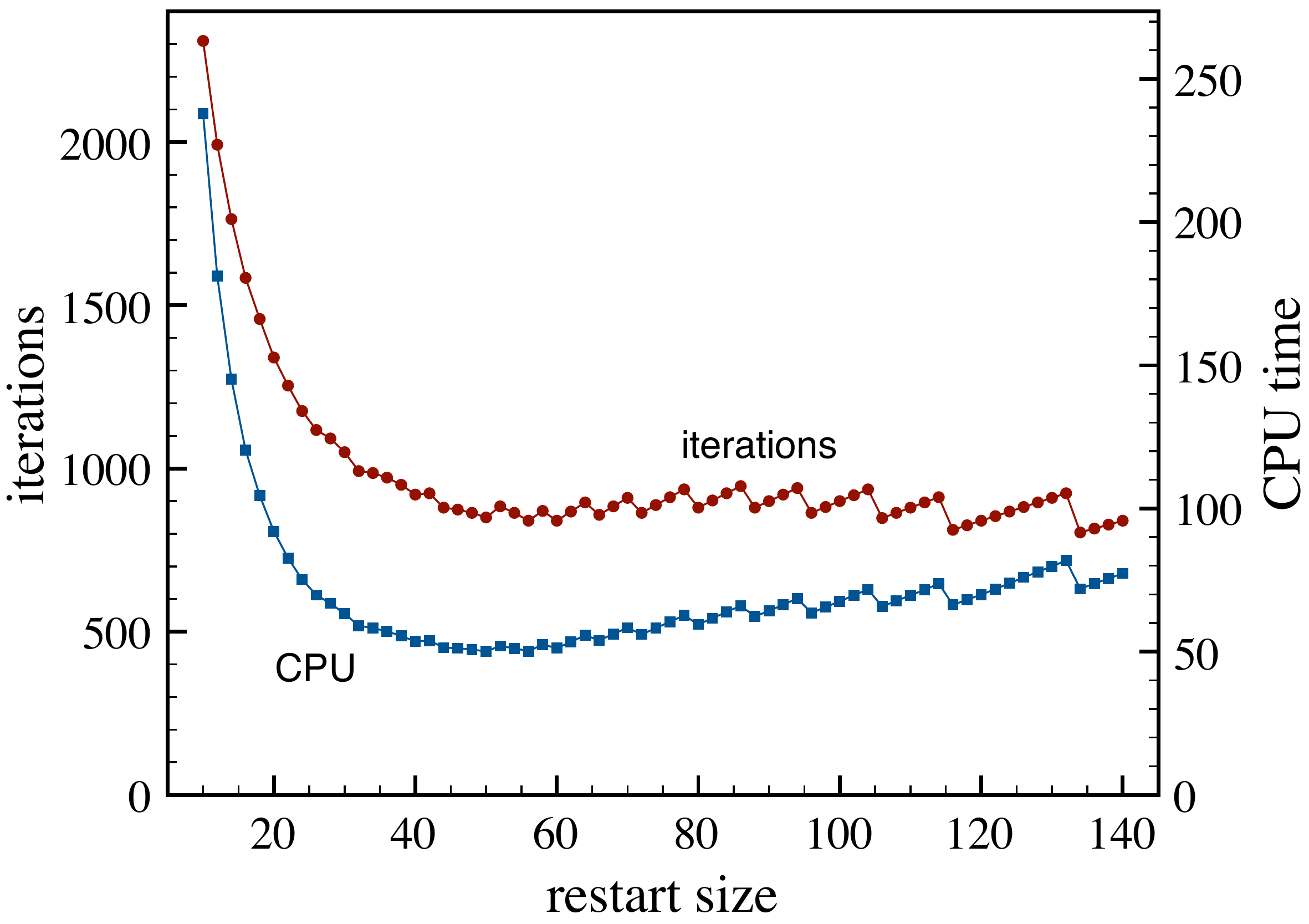}}
\caption{\label{Fig:restart}Dependence of the number of iterations and
  run time (in seconds) on the restart size for the restarted
  FOM-LR method to achieve an accuracy of $10^{-8}$ for an $8^4$
  configuration (left) and a $10^4$ configuration (right).  The
  deflation gap is 0.1 in both cases.\vspace{-8pt}}
\end{figure}

The restarted FOM-LR method contains three tunable parameters: the
deflation gap $m$, the number $s$ of poles in the partial fraction
expansion and the restart size $k_\text{max}$, i.e., the maximal size
of the Krylov subspace before restarting.  Figure~\ref{Fig:restart}
shows the effect of the restart size on the CPU time used by the
restarted FOM-LR method. Clearly, there is an optimal size which
should be determined before performing production runs.  The number of
poles in the partial fraction expansion is chosen adequately to
achieve the desired accuracy, and strongly depends on the deflation
gap. In our numerical results the number of poles varied between 8 and
70.

\section{Conclusions}
\label{sec:concl}
At nonzero chemical potential, the overlap Dirac operator contains
the sign function of the Wilson operator $H_W =\gamma_5 D_W$, which is
non-Hermitian.  The by far most expensive part when applying the
overlap Dirac operator to a field vector $b$ --- the standard step in
any iterative solver for the overlap Dirac operator --- is the
computation of the action
of $\sign(H_W)$ on $b$. As a
step towards developing computationally feasible methods for the
dynamical simulation of overlap fermions at nonzero chemical
potential, we proposed in this paper several short-recurrence Krylov
subspace methods to efficiently compute $\sign(H_W)b$.

One class of methods is based on restarts of the Arnoldi process and
requires a precise rational approximation for the sign function on the
(complex) spectrum of the Wilson operator. This means that we need to
have information on the location of the spectrum in the complex plane
and that we have to adapt the number of poles in the rational
approximation accordingly.  The storage requirements for these methods
depend on the restart value, a parameter which has to be tuned to
be optimal, and the number of poles in the rational approximation.
Storage does not depend on the number of iterations to be performed. 

The other class of methods relies on the two-sided Lanczos process. 
We can use a rational function approximation, in which case the comments made in the previous paragraph apply as well, except that there is no restart.
Alternatively, the sign function can be evaluated directly. In that case,
if a two-pass strategy is used, the storage requirements are minimal;
otherwise storage increases linearly with the number of iterations. No
\emph{a priori} knowledge on the spectrum is required. If the number
of iterations to be performed gets large, the work spent in evaluating
the sign function of the projected operator, which is represented by a
tridiagonal matrix, becomes decisive in terms of computational cost.
Therefore, the methods based on a rational function approximation were
faster in the numerical experiments that we performed on lattices with
sizes ranging from $4^4$ to $10^4$. However, fast methods are
currently being developed to compute the sign of the projected
tridiagonal matrix, which will speed up the direct two-sided Lanczos 
method substantially \cite{Bloch:2009uc}.

For both classes of methods the deflation of critical eigenvalues is
an important ingredient towards efficiency. We showed that LR
deflation is to be preferred to Schur deflation.

\section*{Acknowledgements} 

We would like to thank R\'emy Lopez, who during his internship at the
University of Wuppertal worked out the C implementation of the Arnoldi
based methods.

\bibliographystyle{elsarticle-num}
\bibliography{my_lit}

\begin{thebibliography}{10}
\expandafter\ifx\csname url\endcsname\relax
  \def\url#1{\texttt{#1}}\fi
\expandafter\ifx\csname urlprefix\endcsname\relax\def\urlprefix{URL }\fi
\expandafter\ifx\csname href\endcsname\relax
  \def\href#1#2{#2} \def\path#1{#1}\fi

\bibitem{Narayanan:1994gw}
R.~Narayanan, H.~Neuberger, {A construction of lattice chiral gauge theories},
  Nucl. Phys. B443 (1995) 305--385.
\newblock \href {http://arxiv.org/abs/hep-th/9411108}
  {\path{arXiv:hep-th/9411108}}.

\bibitem{Neuberger:1997fp}
H.~Neuberger, {Exactly massless quarks on the lattice}, Phys. Lett. B417 (1998)
  141--144.
\newblock \href {http://arxiv.org/abs/hep-lat/9707022}
  {\path{arXiv:hep-lat/9707022}}.

\bibitem{Ginsparg:1981bj}
P.~H. Ginsparg, K.~G. Wilson, {A remnant of chiral symmetry on the lattice},
  Phys. Rev. D25 (1982) 2649.

\bibitem{Luscher:1998pqa}
M.~L{\"u}scher, {Exact chiral symmetry on the lattice and the Ginsparg-Wilson
  relation}, Phys. Lett. B428 (1998) 342--345.
\newblock \href {http://arxiv.org/abs/hep-lat/9802011}
  {\path{arXiv:hep-lat/9802011}}.

\bibitem{Bloch:2006cd}
J.~C.~R. Bloch, T.~Wettig, {Overlap Dirac operator at nonzero chemical
  potential and random matrix theory}, Phys. Rev. Lett. 97 (2006) 012003.
\newblock \href {http://arxiv.org/abs/hep-lat/0604020}
  {\path{arXiv:hep-lat/0604020}}.

\bibitem{vandenEshof:2002ms}
J.~van~den Eshof, A.~Frommer, T.~Lippert, K.~Schilling, H.~A. van~der Vorst,
  {Numerical methods for the QCD overlap operator. I: Sign-function and error
  bounds}, Comput. Phys. Commun. 146 (2002) 203--224.
\newblock \href {http://arxiv.org/abs/hep-lat/0202025}
  {\path{arXiv:hep-lat/0202025}}.

\bibitem{Bloch:2007xi}
J.~C.~R. Bloch, T.~Wettig, {Domain-wall and overlap fermions at nonzero quark
  chemical potential}, Phys. Rev. D76 (2007) 114511.
\newblock \href {http://arxiv.org/abs/0709.4630} {\path{arXiv:0709.4630}}.

\bibitem{Bloch:2007aw}
J.~C.~R. Bloch, A.~Frommer, B.~Lang, T.~Wettig, {An iterative method to compute
  the sign function of a non- Hermitian matrix and its application to the
  overlap Dirac operator at nonzero chemical potential}, Comput. Phys. Commun.
  177 (2007) 933--943.
\newblock \href {http://arxiv.org/abs/0704.3486} {\path{arXiv:0704.3486}}.

\bibitem{higham08}
N.~J. Higham, {Functions of Matrices: Theory and Computation}, {Society for
  Industrial and Applied Mathematics}, 2008.

\bibitem{frommer06}
A.~Frommer, V.~Simoncini, Matrix Functions, Vol.~13 of Mathematics in Industry,
  Springer, Heidelberg, 2008, Ch.~3, pp. 275--303.

\bibitem{vdV87}
H.~van~der Vorst, {An iterative solution method for solving {$f(A)x=b$}, using
  {Krylov} subspace information obtained for the symmetric positive definite
  matrix {A}.}, J. Comput. Appl. Math. 18 (1987) 249--263.

\bibitem{Bloch:2007jg}
J.~C.~R. Bloch, T.~Wettig, A.~Frommer, B.~Lang, {An iterative method to compute
  the overlap Dirac operator at nonzero chemical potential}, PoS LAT2007 (2007)
  169.
\newblock \href {http://arxiv.org/abs/0710.0341} {\path{arXiv:0710.0341}}.

\bibitem{eiermann06}
M.~Eiermann, O.~Ernst, {A restarted Krylov subspace method for the evaluation
  of matrix functions}, SIAM J. Numer. Anal. 44 (2006) 2481--2504.

\bibitem{Parlett.80}
B.~N. Parlett, {A new look at the {Lanczos} algorithm for solving symmetric
  systems of linear equations}, Linear Algebra Appl. 29 (1980) 323--346.

\bibitem{Paige.Parlett.vdVorst.95}
C.~C. Paige, B.~N. Parlett, H.~A. {van~der~Vorst}, Approximate solutions and
  eigenvalue bounds from {Krylov} subspaces, Numer. Linear Algebra Appl. 2~(2)
  (1995) 115--134.

\bibitem{saad96}
Y.~Saad, {Iterative Methods for Sparse Linear Systems}, 2nd Edition, SIAM,
  Philadelphia, 2003.

\bibitem{frommer03}
A.~Frommer, {BiCGStab(l) for families of shifted linear systems}, Computing
  70~(2) (2003) 87--109.

\bibitem{simoncini03}
V.~Simoncini, {Restarted full orthogonalization method for shifted linear
  systems}, BIT Numerical Mathematics 43 (2003) 459--466.

\bibitem{frommer98}
A.~Frommer, U.~Gl\"assner, {Restarted GMRES for shifted linear systems}, SIAM
  J. Sci. Comput. 19 (1998) 15--26.

\bibitem{Schaefer:2008phd}
K.~Sch\"afer, Krylov subspace methods for shifted unitary matrices and
  eigenvalue deflation applied to the {N}euberger {O}perator and the matrix
  sign function, Ph.D. thesis, University of Wuppertal (2008).

\bibitem{Bloch:2008gh}
J.~C.~R. Bloch, T.~Breu, T.~Wettig, Comparing iterative methods to compute the
  overlap {D}irac operator at nonzero chemical potential, PoS LATTICE2008
  (2008) 027.
\newblock \href {http://arxiv.org/abs/0810.4228} {\path{arXiv:0810.4228}}.

\bibitem{Jegerlehner:1996pm}
B.~Jegerlehner, {Krylov space solvers for shifted linear systems} (1996).
\newblock \href {http://arxiv.org/abs/hep-lat/9612014}
  {\path{arXiv:hep-lat/9612014}}.

\bibitem{VdVorst:1992}
H.~A. van~der Vorst, {BI-CGSTAB: a fast and smoothly converging variant of
  BI-CG for the solution of nonsymmetric linear systems}, SIAM J. Sci. Stat.
  Comput. 13~(2) (1992) 631--644.

\bibitem{freund92}
R.~W. Freund, N.~M. Nachtigal, {QMR: a Quasi-Minimal Residual Method for
  Non-Hermitian Linear Systems}, Numer. Math. 60 (1991) 315--339.

\bibitem{freund93}
R.~W. Freund, {Solution of Shifted Linear Systems by Quasi-Minimal Residual
  Iterations}, in: L.~Reichel, A.~Ruttan, R.~S. Varga (Eds.), Numerical Linear
  Algebra, W. de Gruyter, 1993, pp. 101--121.

\bibitem{Bloch:2009uc}
J.~C.~R. Bloch, S.~Heybrock, {A nested Krylov subspace method to compute the
  sign function of large complex matrices} (2009).
\newblock \href {http://arxiv.org/abs/0912.4457} {\path{arXiv:0912.4457}}.

\bibitem{golub}
G.~H. Golub, C.~F. van Loan, {Matrix Computations}, 3rd Edition, {Johns Hopkins
  University Press}, 1996.

\bibitem{zolotarev77}
E.~I. Zolotarev, {Application of elliptic functions to the question of
  functions deviating least and most from zero}, Zap. Imp. Akad. Nauk. St.
  Petersburg 30 (1877) \!.

\bibitem{ingerman00}
D.~Ingerman, V.~Druskin, L.~Knizhnerman, {Optimal finite difference grids and
  rational approximations of the square root. I. Elliptic problems}, Comm. Pure
  Appl. Math. 53~(8) (2000) 1039--1066.

\bibitem{KL}
C.~Kenney, A.~Laub, A hyperbolic tangent identity and the geometry of {Pad\'e}
  sign function iterations, Numer. Algorithms 7~(2-4) (1994) 111--128.

\bibitem{Neuberger:1998my}
H.~Neuberger, {A practical implementation of the overlap Dirac operator}, Phys.
  Rev. Lett. 81 (1998) 4060--4062.
\newblock \href {http://arxiv.org/abs/hep-lat/9806025}
  {\path{arXiv:hep-lat/9806025}}.

\bibitem{Neuberger:1999zk}
H.~Neuberger, {The overlap Dirac operator}, in: A.~Frommer, T.~Lippert,
  B.~Medeke, K.~Schilling (Eds.), Numerical challenges in Lattice Quantum
  Chromodynamics, Springer Berlin, 2000, pp. 1--17.
\newblock \href {http://arxiv.org/abs/hep-lat/9910040}
  {\path{arXiv:hep-lat/9910040}}.

\end{thebibliography}

\end{document}